\documentclass[floatfix,%
 aip,
 amsmath,amssymb,
 reprint,%
]{revtex4-1}

\usepackage{graphicx}
\usepackage{booktabs}
\usepackage{multirow}
\usepackage{bm}
\usepackage{xcolor}
\usepackage{hyperref}

\newcommand{\benchname}{FlowReconBench}

\begin{document}

\title{\benchname{}: Accuracy, calibration and cost of learned and classical flow reconstruction from sparse sensors}

\author{Nicolas Tricard}
\author{Linzheng Wang}
\author{Jason Chen}
\author{Sili Deng}
\email[Author to whom correspondence should be addressed: ]{silideng@mit.edu}
\affiliation{Department of Mechanical Engineering, Massachusetts Institute of
Technology, Cambridge, Massachusetts 02139, USA}

\date{November 4, 2026}

\begin{abstract}
Reconstructing full flow fields from sparse sensors requires prior assumptions about unmeasured regions and physical variables. Yet it remains unclear when learned models outperform classical estimators and which aspects of their predictions can be trusted.
\benchname{} evaluates ten learned and five classical methods on three regimes: 
a laminar cylinder
wake across Reynolds numbers, two-dimensional Kolmogorov turbulence and
a large-scale three-dimensional (3D) isotropic turbulence. Methods share 
leakage-controlled splits and sensor draws and are scored on per-channel 
error, ensemble spread, calibration, spectra and cost. The leading method changes for each regime. Classical methods perform strongly in 
low-order and dense-sensing regimes. On the cylinder wake, linear stochastic 
estimation and gappy proper orthogonal decomposition lead, including for 
unobserved pressure. In 3D, inverse-distance weighting matches the best 
learned methods on observed channels.
Data-driven methods become superior at decreasing 
sensor density or poor sensor observability. On Kolmogorov flow, continuous ranked
probability score and error yield different method rankings: 
generative models offer broader but underdispersed ensembles, while 
deterministic models prioritize pointwise error.
Generative models balance this tradeoff through their sampler budget, 
with a larger budget improving ensemble quality at higher computational cost.
In 3D, scaling tests demonstrate that queryable methods are flat in memory, while
full-grid construction methods run out of memory. Observing only 2 of 4 fields 
leaves the problem kinematically underconstrained, yet a latent model can reproduce structural similarities to the ground truth. 
\end{abstract}

\maketitle
\raggedbottom

\section{Introduction}\label{sec:intro}

Reconstructing full flow fields from sparse observations can reduce 
experimental and computational costs across a broad range of fluid dynamics
applications.\cite{brunton2020machine,fukami2023super}
In experimental settings, techniques such as hot-wire anemometry\cite{comtebellot1976hot} 
and pressure measurements in aircraft flight testing\cite{whitmore1995rtfads} provide information 
at a limited number of spatial locations, motivating methods capable of recovering the 
surrounding flow structure without requiring dense instrumentation.
\cite{vinuesa2023transformative,mousavi2025low} Similarly, in computational fluid dynamics, 
reconstructing complex flow fields from a small number of sampled 
locations can reduce data storage requirements and enable efficient access 
to flow information without retaining the complete simulation output.\cite{glaws2020deep,
STAHL2025114037} Beyond recovering unmeasured spatial regions, reconstruction may 
also involve inferring unobserved fields, the physical quantities that no sensor 
measures, such as estimating pressure fields from velocity measurements.
\cite{vanoudheusden2013piv} However, sparse observations constrain only a 
small fraction of the degrees of freedom of the full
multivariate flow field, so the set of fields consistent with the measurements
remains large.\cite{zaki2021limited,stuart2010inverse,callaham2019robust,
raissi2020hidden,vinuesa2023transformative}
The reconstruction problem is therefore ill-posed: its solution is non-unique and
sensitive to measurement noise. Recovering the underlying flow structure requires 
prior information that restricts this set to plausible fields.
The central question is therefore which aspects of a flow can be inferred 
from the available measurements, which require a learned prior, and under 
what conditions a more
complex reconstruction method offers a meaningful advantage.

Data-driven methods learn this prior from representative flow
fields,\cite{everson1995karhunen,manohar2018data} encoding spatial correlations,
coherent structures, and relationships between physical quantities as
constraints, and require no knowledge of or solution to the underlying 
governing equations.\cite{berkooz1993proper,taira2017modal,boree2003extended}
Classical methods use the simplest priors. Gappy proper orthogonal decomposition
(POD)\cite{everson1995karhunen,willcox2006unsteady} fits the coefficients of a linear
low-rank basis to the readings, which is often successful for low-rank systems 
but fails when more complex flow structures are present or the sensing is inadequate.
Learned methods replace these with nonlinear priors and differ along three design
choices. The first is the field representation.
The field may be represented in a compressed latent space, as in latent
diffusion,\cite{rombach2022high,du2024confild} 
or directly in ambient space with a fixed grid\cite{li2024s3gm,shu2023physics,li2021fourier}
or at arbitrary query points.\cite{santos2023senseiver,lu2021learning,
kerrigan2024functional,lim2025score}
The second is sensor conditioning. Readings may be rasterized onto a grid with an
observation mask, copied to every grid point from the nearest
sensor,\cite{fukami2021global} gathered from the nearest sensors of each query
point,\cite{wang2026dmfgen} encoded as a set into a fixed-size summary, as in the
attention-based Senseiver,\cite{santos2023senseiver,jaegle2021perceiver} or 
imposed only at sampling time through posterior-guided diffusion methods, 
such as diffusion posterior sampling (DPS) with a pretrained unconditional
prior.\cite{li2024s3gm,amoros2026guiding,chung2023diffusion}
The third is inference. Deterministic regressors
\cite{erichson2020shallow,santos2023senseiver,dang2025flronet,zhang2025operator}
trained on squared error estimate the conditional mean, which suppresses the 
fine-scale fluctuations between the sparse sensors.\cite{oommen2026turbulent,
yao2026posteriorbenchpointestimatesposterior,
zhou2026physicalfieldreconstructionsparseobservations}
Generative models based on diffusion
\cite{li2024s3gm,shu2023physics,huang2024diffusionpde,amoros2026guiding,oommen2026turbulent}
or flow matching\cite{lipman2023flow,liu2023flow} instead sample approximately 
from the distribution of fields consistent with the readings, reflecting the 
non-uniqueness of the inverse problem.
Together these choices decide which possible discretizations and sensor and query locations a
method can be applied to,\cite{kovachki2023neural,fukami2021global} and whether it is able to infer
unobserved channels.\cite{guastoni2021convolutional,cuellar2024three,Wang2026Learning}
They also combine nearly freely. One Fourier backbone can serve as a deterministic
regressor or as the network of a flow-matching model, and a latent prior can be
conditioned during training or through posterior guidance. Published methods therefore
differ in several choices at once.

%
Popular benchmarks for scientific machine learning method comparison, such as
PDEBench,\cite{takamoto2022pdebench} The Well,\cite{ohana2024well}
CFDBench\cite{luo2023cfdbench} and AirfRANS,\cite{bonnet2022airfrans} pose
forward-surrogate tasks and, in PDEBench, inverse problems, but none poses
reconstruction from scattered point sensors. Their datasets are nonetheless used for
sparse reconstruction,\cite{dang2025flronet,huang2026fluidsplat} with splits, sensor
layouts and metrics chosen separately by each study. 
DynaBench\cite{dulny2023dynabench} and StocBench\cite{pfister2026stocbench} pose
forecasting rather than reconstruction, from scattered low-resolution measurements and
for a stochastically forced two-dimensional (2D) Kolmogorov flow, respectively. OceanBench\cite{johnson2023oceanbench}
poses sparse reconstruction but is restricted to sea surface height from simulated
satellite observations.
Sparse sensing benchmarks especially incorporating modern generative techniques are rare.
Two concurrent works, the study of Zhou et
al.\cite{zhou2026physicalfieldreconstructionsparseobservations} and 
PosteriorBench,\cite{yao2026posteriorbenchpointestimatesposterior} assess the
advantages of generative modeling and its calibration, respectively, but focus on the
inference strategy. Neither compares against classical interpolation or gappy POD or tests three-dimensional (3D)
turbulence. Overall, several factors remain under-studied:

\begin{enumerate}
\item \emph{Architectural trade-offs are poorly characterized.} Beyond the deterministic versus 
generative tradeoff, sparse reconstruction methods employ different field representations and 
sensor-conditioning mechanisms, yet their consequences for 
domain compatibility, reconstruction accuracy, physical fidelity, and computational scalability 
are rarely examined together.
\item \emph{Random frame splits leak test information.} Adjacent snapshots of a
simulation are strongly correlated, yet several studies split snapshots at
random\cite{santos2023senseiver,erichson2020shallow,amoros2026guiding} or test on later
frames of a training trajectory,\cite{zhang2025operator} which places near-copies of
the test frames in training.
\item \emph{Ensembles are rarely scored.} Generative reconstruction
papers usually show mean and standard-deviation maps. Apart from recent
work,\cite{zhou2026physicalfieldreconstructionsparseobservations,yao2026posteriorbenchpointestimatesposterior} 
few report a proper scoring rule\cite{gneiting2007strictly} or a calibration diagnostic for
single-snapshot reconstruction from scattered sensors, and the scored ensembles are
often under-dispersed.
\cite{rybchuk2023ensemble,manshausen2025generative,zhou2026physicalfieldreconstructionsparseobservations,yao2026posteriorbenchpointestimatesposterior}
\item \emph{Classical baselines are missing.} Interpolation, such as
inverse-distance weighting (IDW),\cite{shepard1968idw} and gappy POD are rarely
reported; several studies test learned methods on a single periodic cylinder wake
which can be adequately solved using these methods.\cite{santos2023senseiver,zhang2025operator}
\end{enumerate}

Here we introduce \benchname{} to address these gaps with three regimes of increasing 
dimension and dynamical complexity (Sec.~\ref{sec:datasets},
Table~\ref{tab:regimes}): laminar cylinder vortex shedding across Reynolds
numbers, where low-rank linear methods should be strong, with a sensing variant that
reconstructs the full field from body-surface taps; two-dimensional
Kolmogorov turbulence, a standard testbed of generative
reconstruction;\cite{shu2023physics,li2024s3gm} and dense three-dimensional isotropic
turbulence from the Johns Hopkins turbulence database (JHTDB),~\cite{li2008public} 
where the sensors underconstrain by observing two of four channels.
Each regime uses a different holdout, channel structure and resolution. 
We test fifteen methods, ten learned and five classical, using shared dataset
splits and sensor draws where possible, and each is
scored on per-channel accuracy, uncertainty calibration, turbulence spectra, 
and training and inference computational cost. The ten learned
methods range from deterministic operator learners to latent and ambient
generative models and are matched at 6.5M parameters where the published
configurations allow (Sec.~\ref{sec:methods}). The main contributions are an 
implementation of a vast array of methods;
a fair training protocol where training and test sets are not obtained through
shuffling timesteps from a transient simulation; probabilistic scoring, which shows that most ensembles 
are under-dispersed at error-minimizing sampler settings; two diagnostics 
that explain why classical methods are strong; and data and code, which will be
released upon publication.

Section~\ref{sec:design} describes the benchmark and Section~\ref{sec:methods}
the methods.
Section~\ref{sec:results} reports each regime. Section~\ref{sec:analysis}
compares the regimes. Section~\ref{sec:discussion} 
turns these results into constraints and guidelines for choosing a method.
\section{Benchmark design}\label{sec:design}

\subsection{Task}\label{sec:task}

Let $u:\Omega\to\mathbb{R}^{C}$ be a flow field with $C$ channels. Sensors
return readings $\mathcal{O}=\{(\bm{x}_j,c_j,y_j)\}_{j=1}^{M}$ with
$y_j=u_{c_j}(\bm{x}_j)+\varepsilon_j$, where $\bm{x}_j$ is the sensor location,
$c_j$ the observed channel and $\varepsilon_j$ the noise. A method estimates $u$
at all query locations from $\mathcal{O}$ alone or, if it is probabilistic,
samples an approximation of $p(u\mid\mathcal{O})$. It has no access to sensor
time histories, initial conditions or the governing equations. Sensors may be
scattered or confined to a surface, and they may observe only some of the
channels.

\subsection{Datasets: three flow regimes}\label{sec:datasets}

\begin{table*}[tbp]
\caption{\label{tab:regimes}The three regimes. The split column names the
leakage control (Sec.~\ref{sec:datasets}).}
\footnotesize
\begin{ruledtabular}
\begin{tabular}{llllll}
Regime & Dataset & Points & Channels & Split & Primary challenge \\
\hline
2D laminar periodic & Cylinder wake (new) & $2.4{\times}10^{4}$ & $u_x,u_y,p$ & cross-Re & Low-rank observability \\
2D chaotic & Kolmogorov\cite{shu2023physics} & $6.6{\times}10^{4}$ & $\omega$ & cross-traj. & Multiscale reconstruction \\
3D chaotic & JHTDB isotropic\cite{li2008public} & $1.95{\times}10^{6}$ & $U_x,U_y,U_z,p$ & spatial & Unobserved channels \\
\end{tabular}
\end{ruledtabular}
\end{table*}

The three datasets test different kinds of generalization: unseen Reynolds
numbers of a periodic wake, new trajectories of a statistically stationary
flow, and a separated region of the same direct numerical simulation at the
same instants.

\paragraph{Cylinder wake.} Two-dimensional incompressible flow past a circular
cylinder was computed with OpenFOAM~9\cite{weller1998tensorial,openfoam9} 
(\texttt{pimpleFoam}, laminar) on a 12-block O-grid of 23{,}800 cells 
at Reynolds numbers $\mathrm{Re}\in\{60,80,100,150,200,250\}$
($D=U_\infty=1$). After a transient of 120 time units, each case is sampled
every 0.5 units, giving 300 snapshots, or about 25 shedding cycles at
$\mathrm{Re}=100$. The Strouhal numbers (0.143 to 0.206), from the period of
the lift coefficient, fall 1\%--6\% from measured
values\cite{williamson1996vortex,norberg2003fluctuating}. Models train on
$\mathrm{Re}\in\{60,100,150,200\}$ and are tested at $\mathrm{Re}=80$, an
interpolation, and $\mathrm{Re}=250$, an extrapolation. Sensors observe $u_x$
and $u_y$ and pressure $p$ is unobserved. The 360 cells whose centers lie
within $0.03D$ of the wall form the sensor pool of the surface task 
(Sect.~\ref{sec:surface}).

\paragraph{Kolmogorov turbulence.} The two-dimensional forced-turbulence dataset
of Ref.~\onlinecite{shu2023physics} is a common testbed for generative
reconstruction.\cite{amoros2026guiding} It holds vorticity on the doubly
periodic box $[0,2\pi)^2$ at Reynolds number 1000 with wavenumber-4 forcing, in
40 trajectories of 320 frames at $256^2$. Every fourth frame is kept (80 per
trajectory, 3{,}200 in total). Whole trajectories are held out: 32 train and 8
evaluate, assigned by a fixed seeded permutation.

\paragraph{3D isotropic turbulence.} Four $125^3$ cubes at disjoint locations
are cut from the JHTDB \texttt{isotropic1024coarse} direct numerical simulation
(DNS)\cite{li2008public}
(forced isotropic turbulence, Taylor-microscale Reynolds number $Re_\lambda\approx418$). Each cube spans about
0.56 integral lengths at a grid spacing of about 2.1 Kolmogorov lengths. Fifty
snapshots, 100 database steps apart, store $(U_x,U_y,U_z,p)$, standardized per
channel. Cubes 1--3 train, and the spatially disjoint cube 4 is held out; the
nearest training cube lies 0.76 integral lengths from it.
Sensors observe $U_x$ and $U_z$; $U_y$ and $p$ are never observed.

The common approach of shuffling the snapshots of one simulation between training
and test sets\cite{santos2023senseiver,erichson2020shallow,amoros2026guiding} is leaky.
Adjacent stored Kolmogorov frames correlate at $r=0.91$, so a random split in
time would place near-copies of each test frame in training. 
In practice the learned models themselves gain little from seen frames:
re-scored models on frames of training trajectories within $r=0.97$ of a training
frame improves performances by at most 1.2\% beyond an IDW control, except Senseiver
(5.5\%; supplementary material, hereafter SM, Sec.~S6).
On the contrary, under a random frame split these neighbors are still mostly identical 
to training frames, so a sensor-free retrieval would lead the benchmark if it were included. 
Averaging the two stored neighbors of a Kolmogorov test frame, without any sensor, gives 
a relative $L_2$ error [Eq.~(\ref{eq:relL2})] of 0.252, lower than the error of every learned method at 1\% sensing under
the trajectory holdout (Sec.~\ref{sec:kolmogorov}). Thus instead, the suite holds out whole 
Reynolds numbers (cylinder), whole trajectories (Kolmogorov) and a spatially disjoint cube
(JHTDB) to ensure fair evaluation. The tuning and test datasets are obtained by alternating frames of the same held-out data. 
The tuning half selects only a few
discrete settings: checkpoint, sampler steps, guidance weight, raw or averaged weights. Every method uses the same halves, so
the comparison stays fair.

\subsection{Sensor protocols}\label{sec:sensors}

Sensors are drawn uniformly at random over the points of each discretization. On
the cylinder this is uniform over mesh cells, so the sensor density per unit
area is highest near the body, where the O-grid clusters cells. Headline results
use 1\% of the points; sweeps cover 0.1--10\% on Kolmogorov
(Sec.~\ref{sec:density}) and, for the classical anchors only, 0.1--10\% on
JHTDB (SM, Fig.~S1). Training layouts are redrawn
every step; evaluation layouts are frozen and shared by all methods. 
The layout is part of the protocol because sensing configuration plays a dominant 
role on model performance.
On the Kolmogorov test frames, 655 scattered sensors leave
a mean and maximum nearest-sensor distance of 5.1 and 20 grid cells, whereas a
$25{\times}26$ lattice leaves 3.9 and 7.1 cells and lowers the IDW error on every
frame, from 0.545 to 0.451.
Reconstruction from uniformly downsampled fields,\cite{shu2023physics} which are
lattices, is therefore an easier problem than scattered sensing at the same
fraction. Gaussian noise and a slab
occlusion of 25\% of the sensors are tested on Kolmogorov (SM, Sec.~S2).

\subsection{Metrics}\label{sec:metrics}

\paragraph{Accuracy.} The relative $L_2$ error of channel $c$ is
\begin{equation}
  e_c=\frac{\|\hat{u}_c-u_c\|_2}{\|u_c\|_2},
  \label{eq:relL2}
\end{equation}
with fields standardized by training statistics, so the training mean scores 1.
All points weigh equally. On the cylinder mesh this favors the near-body region,
where cells cluster. The aggregate averages the channels. Generative methods report
the error of the ensemble mean and the mean error of single members.

\paragraph{Probabilistic scores.} For $K$ members $s_k$ and truth $y$ at a
point, the continuous ranked probability score (CRPS)\cite{gneiting2007strictly}
is estimated without bias as
\begin{equation}
  \mathrm{CRPS}=\frac{1}{K}\sum_{k=1}^K|s_k-y|-\frac{1}{2K(K-1)}\sum_{k\ne l}|s_k-s_l|
  \label{eq:CRPS}
\end{equation}
and averaged over points. For a deterministic prediction it equals the mean
absolute error (MAE), so deterministic and generative models alike share one scale. The 
spread--error ratio is the root-mean-square (RMS) ensemble standard deviation 
divided by the RMS error of the ensemble mean and
measures whether the ensemble spread matches the size of the actual error.
Coverage$_{90}$ is the fraction of points whose truth lies in the 5--95\% band
of the members and assesses whether the members bracket the truth as often as a
90\% band should. All ensembles have $K=8$ members. A calibrated ensemble of that
size scores about 0.94 and 0.72, not the asymptotic 1 and 0.90, because a finite
ensemble mean carries sampling error and a finite percentile band is
narrow.\cite{fortin2014why}

\paragraph{Physics and cost.} Spectra are computed on single samples of each
method, since averaging removes small scales. Each non-periodic 3D cube has its mean 
removed and is tapered with a variance-preserving Hann window before the fast Fourier
transform; the periodic Kolmogorov fields are not. Cost is in training 
graphics-processing-unit (GPU) hours, peak training memory and inference time per
field on one NVIDIA H100 GPU; 
memory scaling to $1024^3$ points is measured separately (Sec.~\ref{sec:cost}).

\paragraph{Statistical resolution.} Two methods are compared on the same test frames 
and sensor draws. For each frame, the difference of their aggregate errors forms 
one paired sample, giving 50 samples in 2D and 25 on JHTDB. A 95\% confidence 
interval (CI) for the mean difference comes from 20,000 bootstrap resamples of these 
samples. A gap is called resolved only if this interval excludes zero and a 
two-sided Wilcoxon signed-rank test\cite{wilcoxon1945individual} gives $p<0.05$.

\subsection{Benchmark standardization and fairness}\label{sec:fairness}

(i) \emph{Upstream fidelity.} Each external baseline is constructed from its published
code, optimizer, loss and architecture.
Four methods have no upstream template repositories (Sec.~\ref{sec:methods}): latent
flow matching (Latent FM), a multilayer perceptron with a radial-basis-function
kernel (MLP-RBF), a 3D Fourier neural operator (FNO3D) and its flow-matching
variant (FNO3D-FM).
(ii) \emph{Matched budgets.} Every learned method has $6.5$M$\pm10\%$ parameters, 
counting a complex parameter once, except the published prior of the conditional neural field latent diffusion model
(CoNFiLD; 118.9M), which
is constructed for transient-sensing application and thus is tested on approximately
its upstream parameter size. The 2D methods train for about 50k optimizer steps. 
(iii) \emph{Identical conditions.} All methods share data,
splits, normalization, training sensor ranges, sensor draws and metric code,
except the normalization and crop of Gen4Turb (Sec.~\ref{sec:methods}),
and no prediction is overwritten with sensor values.
(iv) \emph{One checkpoint selection rule} (Sec.~\ref{sec:datasets}).

\section{Methods evaluated}\label{sec:methods}
The fifteen methods fall into four groups (Table~\ref{tab:capability}):
classical anchors, deterministic learners, and latent and ambient generative
models.

\begin{table}[tbp]
\caption{\label{tab:capability}Where each method is deployed. Every method run on the
cylinder uses its native mesh; methods that need a Cartesian grid are not run there
(``grid''). ``n.r.'': admitted but not run. det.: deterministic.
Abbreviations (Sec.~\ref{sec:methods}): NN, nearest neighbor; LSE, linear
stochastic estimation; MLP-RBF, multilayer perceptron with a radial-basis-function
kernel; FNO, Fourier neural operator; Geo-FNO, geometry-aware FNO; DeepONet, deep
operator network; FM, flow matching; CoNFiLD, conditional neural field latent
diffusion; S3GM, sparse-sensor-assisted score-based generative model; SiT,
scalable interpolant transformer. $^{\rm a)}$Scored on the central
$120^3$ crop of the held-out cube (Sec.~\ref{sec:methods}).}
\footnotesize
\begin{ruledtabular}
\begin{tabular}{lccc}
Family & Cyl. & Kolm. & JHTDB \\
\hline
Classical (mean, NN, IDW, & \checkmark & \checkmark & \checkmark \\
\quad gappy POD, LSE) & & & \\
Point-attention det.\ (Senseiver) & \checkmark & \checkmark & \checkmark \\
Point-kernel det.\ (MLP-RBF) & \checkmark & \checkmark & n.r. \\
Fourier operator det.\ (Geo-FNO; FNO) & \checkmark & \checkmark & \checkmark \\
Set-DeepONet and Set-DeepONet++ & n.r. & n.r. & \checkmark \\
Latent-grid flow matching (Latent FM) & grid & \checkmark & \checkmark \\
Latent neural-field diffusion (CoNFiLD) & n.r. & n.r. & \checkmark \\
Video diffusion with DPS (S3GM) & grid & \checkmark & \checkmark \\
Voxel diffusion (Gen4Turb) & grid & n.r. & \checkmark$^{\rm a)}$ \\
Fourier flow matching (FNO3D-FM) & grid & n.r. & \checkmark \\
Point-token transformer (SiT-point) & \checkmark & \checkmark & \checkmark \\
\end{tabular}
\end{ruledtabular}
\end{table}

\subsection{Classical anchors} 
Five baselines are tested against the learned methods: the
training mean, nearest-neighbor (NN) assignment, inverse distance weighting (IDW) 
with exponent 1 over the eight nearest sensors, gappy POD,\cite{everson1995karhunen}
and linear stochastic estimation (LSE).\cite{adrian1988stochastic}
Gappy POD fits the coefficients
of a POD basis of the training frames to the sensor readings by least squares.
LSE maps the readings to the full field, unobserved channels included, by ridge
regression over the training frames, refitted for each sensor layout.
Its rank, and the ridge penalty of LSE, are chosen on held-out training frames,
per sensor count in 2D and once on JHTDB; the rank never 
exceeds the number of readings. Neighbor searches are
periodic on Kolmogorov flow.

\subsection{Deterministic learners} 
Senseiver\cite{santos2023senseiver} maps an
arbitrary sensor set to query values through cross-attention into a fixed latent
array. On the periodic Kolmogorov grid, a Fourier neural operator
(FNO)\cite{li2021fourier} reads the rasterized sensors as a value channel and a
mask channel. On the cylinder, Geo-FNO\cite{li2023geofno} runs on the native
mesh. A learned deformation maps the mesh to a latent torus. On JHTDB, FNO3D
reads the rasterized sensors with a 3D FNO and regresses the field directly. 
MLP-RBF, as described in Ref.~\onlinecite{wang2026dmfgen}, encodes each query
with a multilayer perceptron (MLP), gathers sensor tokens for it with a softmax
radial-basis-function (RBF) kernel,\cite{wang2026dmfgen} appends a pooled global summary and is
trained by regression. DeepONet\cite{lu2021learning}
needs readings at a fixed set of locations, which the varying sensor draws do not
provide. Set-DeepONet, run on JHTDB, therefore replaces its branch with a DeepSets
encoder\cite{zaheer2017deep} that averages per-sensor features, a simpler form of
the attention-weighted pooling of the variable-input deep operator network
(VIDON).\cite{prasthofer2022variable}
Set-DeepONet++ replaces this global average by averages over $4^3$ and
$8^3$ spatial bins plus Fourier moments of the readings, and the trunk by random
Fourier features.

\subsection{Latent generative} 
Latent flow matching (Latent FM) is an in-house
implementation similar to latent diffusion:\cite{rombach2022high} a convolutional autoencoder
compresses full fields, and a rectified flow\cite{lipman2023flow,liu2023flow} in
latent space is conditioned on the sensors rasterized as a masked value image.
CoNFiLD\cite{du2024confild} compresses each snapshot with a conditional
sinusoidal representation network (SIREN) autodecoder,~\cite{sitzmann2020siren} trains an unconditional diffusion prior over the
latents, and conditions by DPS.\cite{chung2023diffusion} Although it is 
typically deployed in transient sensing application,
here it is applied on static-snapshots. Its prior generates windows of 32 consecutive
snapshots; only the target snapshot of each window carries sensors, so it receives
the same information as the single-snapshot methods. Its published configuration contains 118.9M
parameters, 113.7M of them in the prior U-Net, and is a declared exception to the
parameter count rule. Parameter-count matching arms are in the SM.

\subsection{Ambient generative} 
Gen4Turb\cite{oommen2026turbulent} is an
ambient generative approach styled after
Ref.~\onlinecite{karras2022elucidatingdesignspacediffusionbased}, consisting of a 3D U-Net
diffusion model conditioned on a random observation mask. Its U-Net needs every
side to be divisible by 8, so it reconstructs the central $120^3$ crop of each
cube. It is trained with its upstream code, loader and min--max normalization on
the three training cubes, with the observation mask restricted to $U_x$ and $U_z$.
It is scored with the shared sensor draw restricted to the crop, keeping 88\%
of both the points and sensors to produce the same sensor density as the other tests. S3GM\cite{li2024s3gm} is an unconditional video-diffusion prior
conditioned at sampling time by DPS. In 3D it treats one spatial axis as video
time. In 2D its prior models single snapshots and is sampled with the upstream
sampler; only the guidance coefficient is chosen on the tuning half, because the
published value diverges on some tuning frames. SiT-point adapts the 
scalable interpolant transformer (SiT)\cite{ma2024sit} with
a point tokenizer: each query point is a token carrying its coordinates, and the
sensor readings reach the tokens by nearest-sensor fill. At sampling time, the 
network sees random subsets of 8,192 query points, the training size. All points
form one state of the sampling ordinary differential equation, and the partition 
is redrawn at every drift evaluation, so
no point is sampled in a fixed subset independently of its neighbors. On the
3D cube a subset holds 0.4\% of the points, and single members still carry
grid-scale speckle (Sec.~\ref{sec:spectra}). FNO3D-FM uses the FNO3D
backbone as the velocity network of a conditional rectified flow,\cite{liu2023flow}
which maps a Gaussian-process sample to the field. Its loss adds a penalty
(weight 0.02) on the mismatch of the binned power in radial wavenumber shells.
It shares the inputs, optimizer and training schedule of FNO3D, so the pair
differs in its objective and this penalty.

\section{Results by regime}\label{sec:results}

Each subsection reports one sparse sensing regime at 1\% random sensing on the 
test data, with shared sensor draws and $K{=}8$ random samples for generative methods. 
While raw results are presented here, Sec.~\ref{sec:analysis} explains the cross-
regime patterns that recur between the tested methods.

\subsection{Laminar vortex shedding}\label{sec:cylinder}

\subsubsection{Volumetric sampling}

Sensors observe $u_x$ and $u_y$ at 238 mesh cells each (1\%). $p$ is unobserved
(Table~\ref{tab:cyl}). Table~\ref{tab:capability} specifies the methods 
not run here due to their requirement of Cartesian grids or otherwise.

\begin{table*}[tbp]
\caption{\label{tab:cyl}Cylinder wake at the held-out
$\mathrm{Re}\in\{80,250\}$ on 25 test frames each using 1\% sensing of $u_x$ and 
$u_y$. Generative methods use $K{=}8$ members and $N$ sampler steps.
Spr/Err: spread--error ratio. Interpolation
returns the training mean on $p$. Boldface denotes the lowest aggregate and
unobserved-pressure errors.}
\footnotesize
\begin{ruledtabular}
\begin{tabular}{lcccccccccc}
 & \multicolumn{6}{c}{Rel.\ $L_2$} & & & & \\
\cline{2-7}
Method & agg. & $\mathrm{Re}=80$ & $\mathrm{Re}=250$ & $u_x$ & $u_y$ & $p$ & CRPS & Spr/Err & Cov.$_{90}$ & s/field \\
\hline
SiT-point ($N{=}8$) & 0.073 & 0.038 & 0.108 & 0.057 & 0.105 & 0.057 & 0.034 & 0.51 & 0.40 & 0.50 \\
Senseiver & 0.134 & 0.118 & 0.150 & 0.115 & 0.192 & 0.095 & 0.082 & --- & --- & 0.06 \\
MLP-RBF & 0.318 & 0.260 & 0.376 & 0.220 & 0.536 & 0.197 & 0.165 & --- & --- & 0.07 \\
Geo-FNO & 0.079 & 0.058 & 0.100 & 0.063 & 0.109 & 0.064 & 0.045 & --- & --- & 0.26 \\
\hline
LSE & \textbf{0.043} & 0.036 & 0.050 & 0.032 & 0.058 & \textbf{0.039} & 0.022 & --- & --- & --- \\
Gappy POD ($r{=}20$) & 0.057 & 0.045 & 0.069 & 0.043 & 0.077 & 0.050 & 0.029 & --- & --- & --- \\
IDW ($k{=}8$) & 0.712 & 0.674 & 0.751 & 0.522 & 0.615 & 1.000 & 0.441 & --- & --- & --- \\
Nearest neighbor & 0.713 & 0.673 & 0.752 & 0.526 & 0.612 & 1.000 & 0.423 & --- & --- & --- \\
Training mean & 1.000 & 1.000 & 1.000 & 1.000 & 1.000 & 1.000 & 0.664 & --- & --- & --- \\

\end{tabular}
\end{ruledtabular}
\end{table*}

The two linear estimators are the most accurate methods. LSE reaches 0.043
relative $L_2$ overall and 0.039 on the unobserved pressure, and gappy POD with 20
modes 0.057 and 0.050. Twenty POD modes hold 99.7\% of the training dataset
energy, and the volume sensors can extract their amplitudes well. 
LSE fits every mode, with a penalty that shrinks the
low-energy ones, instead of truncating the basis (SM, Sec.~S7). On paired frames
LSE leads gappy POD by 0.014 (95\% CI $[-0.016,-0.012]$) and the best learned
method, SiT-point (0.073), by 0.030 ($[-0.040,-0.021]$).

While every learned method recovers the unobserved pressure, none meets the linear
estimators. Section~\ref{sec:density} explains why the classical methods are 
superior here and not elsewhere. Pressure is additionally well-reconstructed by all 
learned models (Sec.~\ref{sec:identifiability}).
SiT-point, the only generative method run on the cylinder, is under-dispersed
relative to the calibrated $K{=}8$ values when pooled over the two Reynolds
numbers, with a spread--error ratio of 0.51 and coverage$_{90}$ of 0.40
(Sec.~\ref{sec:dp}).

Splitting the test frames by Reynolds number separates an interpolated
test-set $\mathrm{Re}=80$ from the extrapolated test trajectory
value $\mathrm{Re}=250$. At the interpolated point, SiT-point nearly catches
LSE (0.038 vs 0.036) and beats gappy POD (0.045).
For the extrapolation, LSE and gappy POD both outperform SiT-point, with Geo-FNO
roughly tied with SiT-point as well. Extrapolation raises the error of 
SiT-point 2.9-fold and that of LSE and gappy POD 1.4- and 1.5-fold (SM, Table~S2).

\begin{figure}[tbp]
\includegraphics[width=\linewidth]{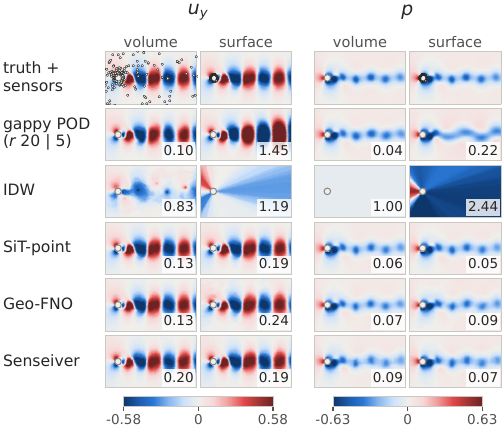}
\caption{\label{fig:qual-cyl}Cylinder wake held-out $\mathrm{Re}=250$ frames
sensed in the volume (238 sensors per field on $u_x$ and $u_y$; $p$ unobserved)
and at the surface (128 taps per field on all three fields). Columns display
$u_y$ and $p$ for each sensing geometry; circles on the truth row mark the 
sensors.  Panel numbers are single-member relative $L_2$, which 
exceeds the ensemble-mean errors of
Tables~\ref{tab:cyl} and~\ref{tab:surface}. Gappy POD is accurate from the volume
sensors and fails from the taps; IDW returns the training mean on the unobserved
pressure.}
\end{figure}

\subsubsection{Surface sensing}\label{sec:surface}

Experiments often measure only on a body surface. The surface task repeats the
cylinder benchmark with sensors confined to the 360 near-wall cells. The
flow, held-out Reynolds numbers, frames and methods are unchanged. Unlike the
volume task, all three fields, including $p$, are read at the taps. The budget
runs from 32 taps per field to the full ring (Table~\ref{tab:surface}), and
generative methods are tuned at 128 taps.

Gappy POD falls from second to last among the non-interpolating methods
(0.594 at 128 taps), because the near-wall taps observe the POD modes poorly
(Sec.~\ref{sec:density}). A ridge penalty lowers its error to 0.262, ahead of
MLP-RBF (0.303) but behind the other three learned methods, and LSE reaches 0.336.
IDW and nearest neighbor exceed 2.0, worse than the training mean, because most
of the wake lies far from every tap.

SiT-point leads at 128 taps, ahead of Senseiver and
Geo-FNO (0.091 versus 0.101 and 0.173). Moving the sensors to the wall raises the 
error of Geo-FNO 2.2-fold while that of SiT-point rises only by 25\%. 
Senseiver's lower error in the surface task may partly reflect the additional pressure observations.
The members of SiT-point remain under-dispersed (spread--error 0.25,
coverage$_{90}$ 0.37).

Every method except LSE is within about 10\% of its full-ring error by 128 taps,
because additional taps resample the same one-dimensional ring. 
Figure~\ref{fig:qual-cyl} compares the two sensing geometries on one frame.
The contrasting behavior of gappy POD between the two sensing configurations illustrates that low-dimensional flow structure alone does not guarantee accurate reconstruction. 
The dominant modes must also be observable from the available measurements.

\begin{table}[tbp]
\caption{\label{tab:surface}Cylinder surface-to-field task: relative $L_2$ of
the full wake from taps on the body, held-out $\mathrm{Re}\in\{80,250\}$, 50
test frames. Columns give taps per field on the near-wall ring; $u_x$, $u_y$
and $p$ are read at the taps. CRPS and the spread--error ratio (Spr/Err) are at
the full ring. The gappy POD rank, and the ridge strength of ridge gappy POD and
LSE, are chosen per budget on training data; ridge gappy POD keeps $r{=}10$.
Boldface denotes the lowest error with the full ring of taps.}
\footnotesize
\begin{ruledtabular}
\begin{tabular}{lcccccc}
 & \multicolumn{4}{c}{Rel.\ $L_2$ at $n$ taps/field} & & \\
\cline{2-5}
Method & 32 & 64 & 128 & 360 & CRPS & Spr/Err \\
\hline
SiT-point & 0.103 & 0.092 & 0.091 & \textbf{0.091} & 0.040 & 0.25 \\
Senseiver & 0.130 & 0.106 & 0.101 & 0.099 & 0.063 & --- \\
Geo-FNO & 0.236 & 0.182 & 0.173 & 0.157 & 0.086 & --- \\
MLP-RBF & 0.370 & 0.322 & 0.303 & 0.290 & 0.156 & --- \\
\hline
Ridge gappy POD & 0.270 & 0.265 & 0.262 & 0.260 & 0.124 & --- \\
LSE & 0.290 & 0.336 & 0.336 & 0.284 & 0.136 & --- \\
Gappy POD ($r{=}5$) & 0.711 & 0.652 & 0.594 & 0.572 & 0.283 & --- \\
IDW ($k{=}8$) & 2.078 & 2.116 & 2.131 & 2.140 & 1.936 & --- \\
Nearest neighbor & 2.161 & 2.154 & 2.146 & 2.144 & 1.939 & --- \\

\end{tabular}
\end{ruledtabular}
\end{table}

\subsection{Two-dimensional Kolmogorov turbulence}\label{sec:kolmogorov}

Sensors observe vorticity at 655 points (1\%). Table~\ref{tab:kolm} lists every
method on the same test frames and sensor draws, and Sec.~\ref{sec:density} gives
the sensor density sweep.

\begin{table*}[tbp]
\caption{\label{tab:kolm}Kolmogorov flow reconstruction.
Sampler budgets, step counts $N$ and the S3GM guidance coefficient are tuning-half
choices. The conditional methods train at 65--6{,}554 sensors per frame. $E_{\rm s}$: 
test-frame median of the single-member energy at $k=32$--128 relative to the truth. 
Inference time and peak memory per field on one H100. per./non-per.: periodic or
non-periodic neighbor search. Boldface denotes the lowest value in each of the
relative $L_2$ and CRPS columns.}
\footnotesize
\begin{ruledtabular}
\begin{tabular}{lcccccccc}
 & \multicolumn{2}{c}{Rel.\ $L_2$} & & & & & & \\
Method & mean & sample & CRPS & Spr/Err & Cov.$_{90}$ & $E_{\rm s}$ & s/field & GB \\
\hline
SiT-point ($N{=}32$) & 0.293 & \textbf{0.310} & 0.146 & 0.37 & 0.42 & 0.27 & 5.00 & 0.22 \\
Latent FM ($N{=}4$) & 0.463 & 0.523 & 0.240 & 0.57 & 0.55 & 0.30 & 0.11 & 0.16 \\
S3GM ($N{=}1000$) & \textbf{0.273} & 0.363 & \textbf{0.110} & 0.95 & 0.75 & 0.93 & 6.80 & 7.30 \\
Senseiver & 0.465 & 0.465 & 0.329 & --- & --- & 0.023 & 0.07 & 0.81 \\
MLP-RBF & 0.500 & 0.500 & 0.337 & --- & --- & 0.13 & 0.11 & 1.36 \\
FNO & 0.380 & 0.380 & 0.257 & --- & --- & 0.039 & 0.24 & 0.28 \\
\hline
LSE & 0.514 & 0.514 & 0.357 & --- & --- & --- & --- & --- \\
IDW ($k{=}8$) (per.) & 0.530 & 0.530 & 0.376 & --- & --- & --- & --- & --- \\
IDW ($k{=}8$) (non-per.) & 0.545 & 0.545 & 0.386 & --- & --- & 0.30 & --- & --- \\
Nearest neighbor (per.) & 0.580 & 0.580 & 0.370 & --- & --- & --- & --- & --- \\
Nearest neighbor (non-per.) & 0.586 & 0.586 & 0.373 & --- & --- & --- & --- & --- \\
Gappy POD ($r{=}160$) & 0.607 & 0.607 & 0.444 & --- & --- & 0.10 & --- & --- \\
Training mean & 1.000 & 1.000 & 0.755 & --- & --- & --- & --- & --- \\

\end{tabular}
\end{ruledtabular}
\end{table*}

S3GM (0.273) is the most accurate method, 0.021 ahead of SiT-point (0.293,
$p=2.5\times10^{-7}$). FNO is the best deterministic method at
0.380, 0.087 behind SiT-point. Every learned method beats IDW (0.530). 

Gappy POD, with rank 160 chosen on a held-out training trajectory, falls behind. 
Two-dimensional turbulence is not low-rank in the POD sense (Sec.~\ref{sec:density}). 
LSE, which keeps every mode, reaches 0.514, between IDW
and MLP-RBF.

Retraining each learned method except SiT-point with two more seeds
gives a pooled standard deviation of 0.0075 on Kolmogorov and 0.0044 on the cylinder.
Gaps within about two standard deviations, such as Latent FM against Senseiver, are
reported as unordered (SM, Sec.~S1).

The ranking is unchanged under Gaussian noise at $\sigma=0.1$ 
and under a slab that removes 25\% of the sensors.
The most accurate methods suffer most at $\sigma=0.3$: the S3GM sampler diverges
on 3 of the 50 test frames and SiT-point L$_2$ rises 39\%. The slab costs S3GM, which
imposes the readings only at sampling time, about as much as SiT-point (1.71 and
1.64 times the clean error).

\begin{figure}[tbp]
\includegraphics[width=\linewidth]{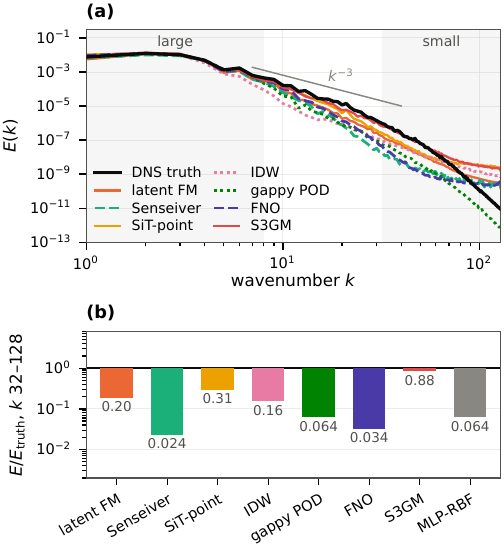}
\caption{\label{fig:spectra-kolm}Kolmogorov spectral fidelity at 655 sensors
(1\%): one held-out frame, one member per method. (a) Kinetic energy spectra. Large scales ($k=1$--8) and small scales
($k=32$--128) are shaded. A curve below the truth at high $k$ is too smooth; a
curve above it carries spurious fine-scale noise. The $k^{-3}$ line is shown for reference. (b) Small-scale band energy
relative to the truth on a log axis centered at 1.}
\end{figure}

\begin{figure}[tbp]
\includegraphics[width=\linewidth]{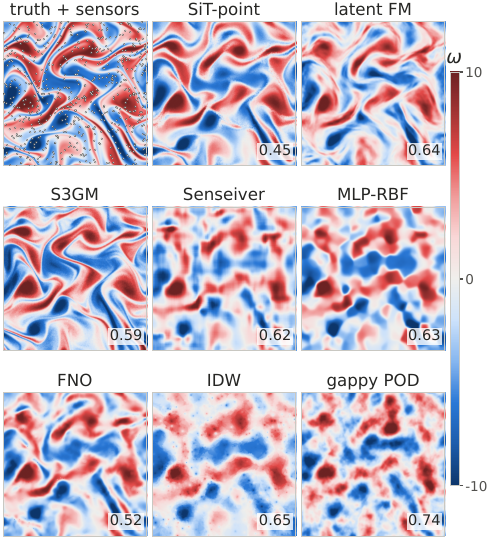}
\caption{\label{fig:qual-kolm}Kolmogorov $256^2$, the held-out frame of
Fig.~\ref{fig:spectra-kolm}: truth and one member per method, with the same 655
sensors (dots on the truth panel) in every panel and one color scale. Panel
numbers are single-member relative $L_2$. Deterministic methods are smooth, most
generative members carry small-scale texture, and gappy POD is over-smoothed.}
\end{figure}

Single members and spectra (Figs.~\ref{fig:spectra-kolm} and~\ref{fig:qual-kolm})
separate the deterministic methods from the generative ones. A regressor trained on
squared error estimates the conditional mean, which averages over all fields
consistent with the sensors. When the sensors leave the small scales
undetermined, this produces a smooth estimate, although each of those fields
carries small-scale structure. Blau and Michaeli\cite{blau2018perception} 
proved that lowering the mean distortion beyond some point forces the distribution of
the estimates away from that of real data. The same tradeoff appears in weather
forecasting\cite{ravuri2021skilful,price2025probabilistic} and in the inpainting
of rotating turbulence, where a generative model does not beat POD pointwise but
reproduces multiscale statistics better.\cite{li2023multiscale} In the
small-scale band $k=32$--$128$, the median test-frame member of Senseiver and FNO
keeps only 0.023 and 0.039 of the true energy, against 0.30 and 0.27
for the generative Latent FM and SiT-point. S3GM comes closest to
the true small-scale energy (0.93).

\subsection{Three-dimensional isotropic turbulence}\label{sec:jhu}

\begin{figure*}[tbp]
\includegraphics[width=\textwidth]{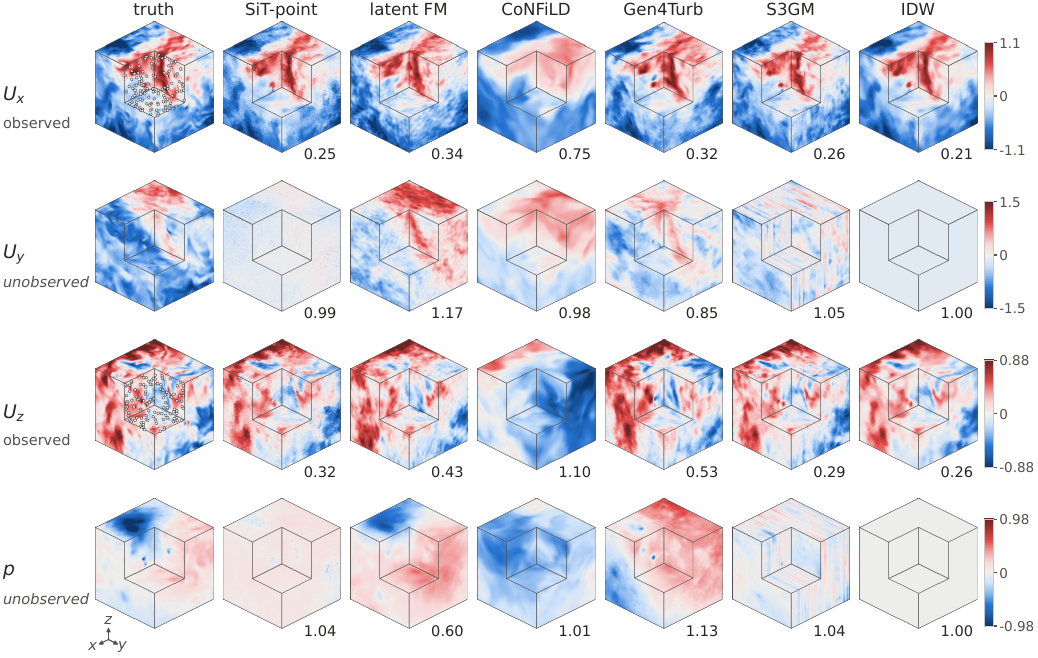}
\caption{\label{fig:qual-jhu}3D isotropic turbulence, held-out cube, frame 152
(a test-half frame). Each panel shows the
$125^3$ cube with one octant cut away to expose the $x$, $y$ and $z$ midplanes.
The other visible surfaces are outer faces of the cube. $U_x$ and $U_z$ are
observed at 19{,}531 identical sensors each (circles: sensors on the midplanes); $U_y$ and
$p$ are unobserved. Gen4Turb shows its central $120^3$ crop. Panel numbers are single-member relative $L_2$ over
the whole cube (the crop for Gen4Turb). Among the methods
shown, only Latent FM is below 0.7 on an unobserved channel, and it identifies the
low-pressure core. IDW returns the training mean on the unobserved channels.}
\end{figure*}

Sensors observe $U_x$ and $U_z$ at 1\% of the $1.95\times10^{6}$ points per
channel. $U_y$ and $p$ are unobserved. On this regular grid every family
except MLP-RBF is run. Table~\ref{tab:jhu} reports accuracy and probabilistic scores on
the held-out cube, Table~\ref{tab:jhucost} the cost, and Fig.~\ref{fig:qual-jhu}
one held-out frame.

\begin{table*}[tbp]
\caption{\label{tab:jhu}3D isotropic turbulence, spatially disjoint held-out
cube ($1.95\times10^{6}$ points). Per-channel relative $L_2$ of the
ensemble mean over the 25 test-half snapshots; agg.\ averages the four channels.
Asterisks mark the unobserved channels, where an error near 1 equals that of the
training mean.
Checkpoints and sampler settings are tuning-half choices. $E_{\rm d}$:
test-snapshot median of the single-member energy at $k=32$--50 relative to the
DNS on the observed channels. Each 3D method is
trained once. Boldface denotes the best relative $L_2$ or CRPS among learned methods.
$^{\rm a)}$Published prior, outside the parameter budget; the 32-snapshot
checkpoints conditioned on sensors in the target snapshot only, with the
guidance scale (0.1) chosen on the tuning half. $^{\rm b)}$Guidance fitted
on the tuning half, because the published coefficients diverge at this size;
final checkpoint of a $\sigma_{\max}{=}20$ prior trained before the frozen protocol.
$^{\rm c)}$Scored on the central
$120^3$ crop (88\% of the points and sensors) and not paired. Further
CoNFiLD, SiT-point and S3GM tests are in SM Table~S3.}
\footnotesize
\begin{ruledtabular}
\begin{tabular}{lccccccccc}
Method & $U_x$ & $U_y^*$ & $U_z$ & $p^*$ & agg. & CRPS & Spr/Err & Cov.$_{90}$ & $E_{\rm d}$ \\
\hline
Latent flow matching & 0.249 & \textbf{0.710} & 0.260 & \textbf{0.643} & \textbf{0.465} & \textbf{0.239} & 0.69 & 0.57 & 0.056 \\
SiT-point ($N{=}32$) & \textbf{0.175} & 0.957 & \textbf{0.182} & 0.904 & 0.555 & 0.323 & 0.46 & 0.40 & 11.7 \\
FNO3D-FM ($N{=}4$) & 0.190 & 0.921 & 0.193 & 1.054 & 0.589 & 0.272 & 0.98 & 0.63 & 0.012 \\
CoNFiLD (single snapshot, $N{=}1000$)$^{\rm a)}$ & 0.500 & 1.024 & 0.543 & 1.034 & 0.775 & 0.378 & 0.77 & 0.62 & --- \\
S3GM (normalized guidance)$^{\rm b)}$ & 0.257 & 0.995 & 0.252 & 0.993 & 0.624 & 0.347 & 0.52 & 0.46 & --- \\
Gen4Turb$^{\rm c)}$ & 0.212 & 0.916 & 0.325 & 0.876 & 0.582 & 0.307 & 0.70 & 0.54 & --- \\
\hline
FNO3D & 0.195 & 0.952 & 0.196 & 0.942 & 0.571 & 0.392 & --- & --- & 0.013 \\
Senseiver & 0.362 & 1.098 & 0.384 & 0.987 & 0.708 & 0.487 & --- & --- & 0.088 \\
Set-DeepONet & 0.527 & 1.098 & 0.567 & 1.056 & 0.812 & 0.583 & --- & --- & --- \\
Set-DeepONet++ & 0.508 & 0.967 & 0.541 & 0.981 & 0.749 & 0.521 & --- & --- & --- \\
\hline
IDW ($k{=}8$, non-periodic) & 0.174 & 1.000 & 0.183 & 1.000 & 0.589 & 0.398 & --- & --- & 1.06 \\
LSE & 0.640 & 1.087 & 0.675 & 1.092 & 0.874 & 0.636 & --- & --- & --- \\
Gappy POD ($r{=}80$) & 0.524 & 1.308 & 0.544 & 1.435 & 0.953 & 0.676 & --- & --- & --- \\
\end{tabular}
\end{ruledtabular}
\end{table*}

\begin{table}[tbp]
\caption{\label{tab:jhucost}Cost of the 3D fleet on one H100-SXM (80\,GB):
parameters, training GPU-hours, peak training memory, and inference time per
full $125^3$ field. (d): derived from evaluation wall time, an upper bound that
includes data loading and metric computation. It marks only slow samplers, so
the orders of magnitude between families are unaffected. $^{\rm a)}$Published CoNFiLD
prior, outside the parameter budget. Optimizer steps and time per step are in SM
Table~S4. Two-stage methods (Latent FM, CoNFiLD) report the sum of both stages and
the larger peak. CPU: run on the central processing unit. RSS: peak resident set
size in host memory.}
\footnotesize
\begin{ruledtabular}
\begin{tabular}{lcccc}
Method & Params & GPU-h & Peak GB & s/field \\
\hline
Latent FM & 6.68M & 20.1 & 8.4 & 0.019 \\
SiT-point ($N{=}32$) & 6.57M & 19.4 & 26.8 & 149 \\
FNO3D-FM ($N{=}4$) & 6.73M & 27.4 & 48.6 & 0.097 \\
CoNFiLD$^{\rm a)}$ & 118.9M & 19.2 & 10.4 & 36 (d) \\
Gen4Turb & 6.16M & 16.6 & 47.4 & 1.65 \\
S3GM & 6.35M & 18.5 & 26.3 & 73.8 \\
Senseiver & 6.42M & 17.0 & 22.4 & $<$1 \\
FNO3D & 6.73M & 24.7 & 48.4 & 0.022 \\
Set-DeepONet & 6.51M & 6.7 & 15.7 & 0.23 \\
Set-DeepONet++ & 6.56M & 1.0 & 10.1 & 0.13 \\
IDW (CPU) & --- & --- & 10 (RSS) & 0.68 \\
Gappy POD (CPU) & --- & --- & 10 (RSS) & 0.14 \\
\end{tabular}
\end{ruledtabular}
\end{table}

While no method leads every column, Latent FM leads the aggregate (0.465), largely
driven by unobserved channels, which it extracts poorly but still better than 
the other methods (0.710 and 0.643). Simultaneously it is the fastest sampler 
of the learned methods. 
While no other method can extract $U_y$ or $p$ (Sec.~\ref{sec:identifiability}),
in the observed channels SiT-point leads the learned methods at 0.175 and 0.182,
level with IDW, which reaches 0.174 and 0.183 (Sec.~\ref{sec:density}). Set-DeepONet's predictions are nearly
constant, a consequence of global mean pooling in its branch. Set-DeepONet++ improves
its aggregate from 0.812 to 0.749 but stays far behind IDW on the observed
channels. FNO3D ties SiT-point on the aggregate (0.571 against 0.555, $p=0.15$).
Gen4Turb, scored on the central crop, errs 1.5 times as much on $U_z$ as on $U_x$
(0.325 against 0.212; SM, Sec.~S4). LSE, fitted to 150 training snapshots of other
cubes, reaches only 0.64 and 0.68 on the observed channels. CoNFiLD errs about
three times as much as IDW on the observed channels (0.500 and 0.543), and
sensors on all 32 snapshots of its window barely change this (0.489 and 0.539;
SM Table~S3). 

At its published guidance coefficients that reproduced upstream outputs
the S3GM sampler diverges at $1.95\times10^{6}$ points.

\section{Cross-Regime Analysis}\label{sec:analysis}

Section~\ref{sec:results} lists raw results within each regime. 
This section discusses which cross-regime 
trends emerge between sensor densities, representations, 
and model families.

\subsection{Pointwise error and CRPS trade off}\label{sec:dp}

Reconstruction error and CRPS assess different properties.
A deterministic regressor trained with squared error estimates the conditional mean,
while CRPS evaluates the quality of the predictive distribution, accounting for both 
accuracy and calibration.
Most models exhibit an inverse trend between error and CRPS. 
When reconstruction uncertainty is high, minimizing the mean-squared error averages over multiple 
plausible solutions, producing blurry estimates, whereas sampling from the posterior 
preserves realistic variability but increases expected error relative to 
the posterior mean.\cite{oommen2026turbulent,blau2018perception} 
In principle, an exact 
posterior sampler could serve both scores. However, in practice,
learned samplers only approximate the posterior, likely reducing mean $L_2$
performance, and with their operating point depending on the sampler, 
the checkpoint, and the number of samples $K$.

Kolmogorov flow shows this tradeoff outside the leader (Table~\ref{tab:kolm}),
S3GM, which leads both scores. Deterministic FNO beats generative Latent FM on
relative $L_2$ (0.380 against 0.463) and loses to it on CRPS (0.257 against
0.240). On JHTDB, the FNO3D pair shares its backbone, inputs and
schedule but differs in objective and in the spectral penalty of FNO3D-FM. The
two tie on relative $L_2$, but the flow-matching model has the lower CRPS.

The tradeoff is not only a difference between architectures but can also emerge
within a given architecture by way of hyperparameter variation. In particular,
one trained model moves along it when only its sampler changes. 
On the Kolmogorov tuning half, raising the step count of Latent FM from 4 to 16
raises its ensemble-mean error from
0.464 to 0.468, lowers its CRPS from 0.240 to 0.226 and raises its spread--error
from 0.57 to 0.83. Few-step members spread less about their mean. Because
sampler settings are selected on the tuning half by ensemble-mean error
(Sec.~\ref{sec:datasets}), the reported few-step methods sit at the low-error end of
their own curves. The settings that minimize CRPS on the tuning half are scored
in SM Table~S5.

At these error-selected settings, the generative models remain largely
under-dispersed against the calibrated $K{=}8$ values (Sec.~\ref{sec:metrics}), a
common observation in related
settings.\cite{rybchuk2023ensemble,manshausen2025generative,
zhou2026physicalfieldreconstructionsparseobservations,
yao2026posteriorbenchpointestimatesposterior} Spread--error ranges from
0.25 (SiT-point on the surface task) to 0.98 (FNO3D-FM on JHTDB), and
Kolmogorov S3GM (0.95) is close to calibrated.

Post-hoc recalibration on held-out data can repair the stated uncertainty
without changing the reconstruction or retraining.\cite{kuleshov2018accurate} A single spread
multiplier fitted on the tuning half, similar to that of Ref.~\onlinecite{anderson2009spatially},
brings all three Kolmogorov generative methods to within 1.5\% of the calibrated
spread--error on the test half. It fails when the miscalibration differs 
within the deployment
set. Fitted separately on the cylinder tuning half, the multiplier for SiT-point
is 1.26 at $\mathrm{Re}=80$ and 3.92 at $\mathrm{Re}=250$.
A correction that changes across the deployment set needs a conditional
estimator fitted separately for each condition. One example is a split-conformal
quantile of the residual over the ensemble spread, fitted in bins of distance to
the nearest sensor (SM, Sec. S7).

\subsection{Performance of classical baselines}\label{sec:density}

Classical approaches offer strong performance and efficiency in a wide variety of 
flow configurations and should not be overlooked as an initial approach to sparse reconstruction.
Several recent studies evaluate learned methods on single periodic cylinder wakes
without classical baselines,\cite{santos2023senseiver,zhang2025operator} and
where interpolation is reported it is competitive.\cite{amoros2026guiding} On
the cylinder task here, which holds out whole Reynolds numbers, LSE and gappy POD
have the lowest pooled relative $L_2$, and interpolation is level with the best
learned method on the observed 3D channels.
This section outlines two descriptors, computed from the training frames and the sensor layout,
that explain why each classical method succeeds or fails: the nearest-sensor
distance relative to the decorrelation length governs interpolation, and the
sparsity and observability of a POD basis govern gappy POD. Together they
locate the regimes in which learned methods outperform both.

Interpolation methods (IDW and nearest-neighbor) can only reuse what nearby sensors 
read, so its error is set by how much the field changes across the gap to the 
nearest sensor. That change is measured by the second-order structure function $S_2(\bm{d})$
\begin{equation}
S_2(\bm{d})=\langle[u(\bm{x}+\bm{d})-u(\bm{x})]^2\rangle~,
\label{eq:variogram}
\end{equation}
\begin{equation}
  \hat{e}_{\mathrm{NN}}=[\langle S_2(\bm{d})\rangle/\langle u^2\rangle]^{1/2}~,
  \label{eq:pred_nn_error}
\end{equation}
where $\bm{d}$ is the distance to the nearest sensor, $u(\bm{x})$ is the field value 
at location $\bm{x}$, $\langle\cdot\rangle$ denotes an average over space and sensor 
draws, and $\hat{e}_{\mathrm{NN}}$ predicts the predicted root-mean-square error of 
the nearest-neighbor variant normalized by the variance $\langle u^2 \rangle$. 
Interpolation error is therefore connected to both the typical sensor spacing 
and the correlation structure of the field.
\cite{matheron1963principles,zimmerman1999experimental,kravchenko2003influence,zhang2015spatial}
Our estimated $\hat{e}$, averaged over random sensor
draws matches the measured one to within 1\% at every Kolmogorov density and to
within 7\% on the cylinder once $S_2$ is resolved in space. It is 8--16\% low on
the 3D cutouts, whose held-out cube has larger velocity increments than the
training cubes (Table~\ref{tab:diagnostics}). 
IDW improves on nearest-neighbor
copying by less than a fifth. Relative to the decorrelation length $\ell$,
defined by $S_2(\ell)=\langle u^2\rangle$, the mean nearest-sensor distance
$\bar d$ at 1\% sensing is below $0.05\,\ell$ for the 3D velocity, $0.4\,\ell$
for Kolmogorov vorticity and about $\ell$ in the cylinder wake, and surface taps
leave the wake more than $10\,\ell$ away. Interpolation is therefore a challenging
baseline to beat on the observed 3D channels, where IDW matches the best
learned methods, a weak one on Kolmogorov, where every learned method beats it at
exponent 1, and worse than the training mean from surface taps.

\begin{table*}[tbp]
\caption{\label{tab:diagnostics}Classical diagnostics at the canonical sensor
counts (test frames). Local columns average the observed channels: $\bar d/\ell$
is the mean nearest-sensor distance over the decorrelation length (cylinder:
wake region $x>D$, $|y|<4D$), $\hat e_{\mathrm{NN}}$ the nearest-neighbor error
predicted from the training structure function and the sensor count
(location-resolved on the cylinder mesh), and NN and IDW the measured errors.
Global columns aggregate all channels: POD rank $r$, training energy fraction
$E_r$, relative $L_2$ of projecting the test frames onto $\Phi_r$ (the floor of
any coefficient fit), median condition number of $H\Phi_r$ over the sensor
draws, and the measured gappy POD error. Kolmogorov local values use the
periodic neighbor search. $^{\rm a)}$In-sample over 150 correlated training
frames.}
\footnotesize
\begin{ruledtabular}
\begin{tabular}{llcccccccccc}
 & & \multicolumn{4}{c}{Local (interpolation)} & & \multicolumn{5}{c}{Global (gappy POD)} \\
\cline{3-6}\cline{8-12}
Regime & Sensors & $\bar d/\ell$ & $\hat e_{\mathrm{NN}}$ & NN & IDW & & $r$ & $E_r$ & Floor & $\kappa(H\Phi_r)$ & Gappy POD \\
\hline
Cylinder, volume & 238 per field & 0.95--1.3 & 0.55 & 0.569 & 0.568 & & 20 & 0.997 & 0.048 & 3.2 & 0.057 \\
Cylinder, surface & 128 taps per field & 13--19 & 2.10 & 2.146 & 2.131 & & 5 & 0.929 & 0.163 & $1.0{\times}10^{2}$ & 0.594 \\
 & & & & & & & 20 & 0.997 & 0.048 & $7.7{\times}10^{3}$ & 0.763 \\
Kolmogorov & 655 & 0.40 & 0.58 & 0.580 & 0.530 & & 160 & 0.832 & 0.513 & 3.7 & 0.607 \\
JHTDB, $U_x$ and $U_z$ & 19{,}531 per channel & $<0.05$ & 0.19 & 0.216 & 0.179 & & 80 & 0.955$^{\rm a)}$ & 0.648 & 11.9 & 0.953 \\
\end{tabular}
\end{ruledtabular}
\end{table*}

A global low-rank method like gappy POD needs two properties. The field must be
\emph{sparse} in some basis: a rank-$r$ POD basis $\Phi_r$ from the training frames must
represent the held-out frames, which the projection floor measures. The retained
modes must also be \emph{observable}: the sampled basis $H\Phi_r$, where $H$
selects the sensed entries, must be well conditioned, or the least-squares fit
amplifies the content outside the basis.
\cite{willcox2006unsteady,peherstorfer2020stability} Volume sensing of the
cylinder meets both conditions (Table~\ref{tab:diagnostics}): 20 modes hold
99.7\% of the training energy, $H\Phi_{20}$ has a condition number of about 3,
and gappy POD (0.057) lands near the floor (0.048). LSE, which keeps every mode
with a penalty weighted by its energy, goes below this rank-20 floor (0.043). 
The surface taps observe the same 20 modes but cannot distinguish them, 
producing a condition number near $8{\times}10^{3}$, so the rank
selection retreats to five modes and still ends at 0.59. Sensing all 360 near-wall
cells does not improve the conditioning.\cite{manohar2018data} A ridge penalty
keeps ten modes and reaches 0.26, still behind three of the four learned methods. Kolmogorov flow and 
the JHTDB cutouts instead fail the compressibility test, with floors of 0.51 and 0.65 
at 160 and 80 modes, respectively. Although their sensors observe the modes well, a 
low-rank field is likewise necessary for gappy POD.

\begin{figure}[tbp]
\includegraphics[width=\linewidth]{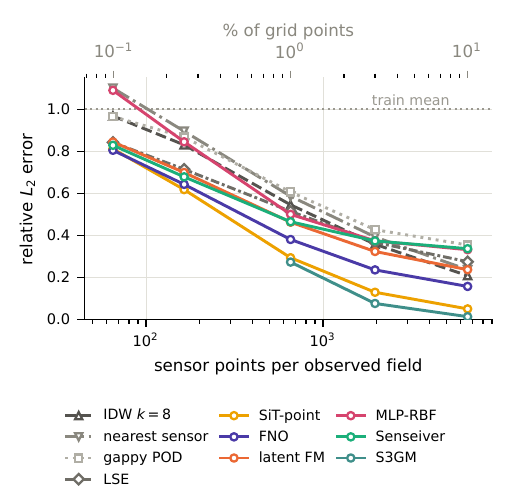}
\caption{\label{fig:sensors}Kolmogorov reconstruction error versus sensors per
field (vorticity, 50 test frames, identical draws for every method at every
density, all densities inside the training ranges). Colored: learned methods; gray:
classical anchors, with the gappy POD rank and the LSE penalty chosen per
density. S3GM scales its guidance weight from the value tuned at 655 sensors
with the inverse square root of the sensor fraction and diverges at 65 and 164
sensors. A 3D sweep of the classical anchors is
in SM Fig.~S1.}
\end{figure}

The Kolmogorov sweep (Fig.~\ref{fig:sensors}) shows that the comparison between
learned and classical methods changes with the sensor budget. At 65 sensors
(0.1\%), IDW (0.968) and gappy POD (0.966) stay near the
training-mean estimate, LSE (0.843) and four learned methods reach
0.805--0.844, MLP-RBF (1.09) exceeds the training mean, and
the S3GM sampler diverges, as it does at 164 sensors.
Denser sensing increasingly favors interpolation. LSE falls behind IDW between 655 and 1965
sensors, and IDW overtakes Senseiver and MLP-RBF in the same interval. It passes
Latent FM by 6554, while S3GM, SiT-point and FNO stay
ahead of IDW (0.013, 0.050, and 0.156 versus 0.208) at 6554 sensors (10\%).

The rate of improvement as a function of sensor count is even more informative. 
The error of S3GM falls 21-fold between 1\% and 10\% sensing alone.
Over the full 100-fold sensor count sweep, the error of SiT-point falls 16-fold,
versus the next best 5.1-fold of FNO. While the two are
0.09 apart at 1\%, they differ by a factor of 3.1 at 10\%. 
The flattest curves belong to models that pass the sensors through a
summary of fixed size or resolution: Senseiver's 64 latent vectors, Latent FM's
$8{\times}8$-cell latent averages and MLP-RBF's fixed-width kernel average. The
steeper curves keep a per-location record of the readings in ambient space
rather than latent space (posterior guidance at the sensor points, nearest-sensor
token fill and a full-resolution sensor raster).
Backbones and training also changing between these models, so the exact
mechanism cannot be isolated.

\subsection{Representation, cross-channel transfer and fine-scale
content}\label{sec:identifiability}\label{sec:spectra}

Methods represent the field either in a compressed latent space, as in latent
diffusion,\cite{rombach2022high} or directly as field values on a grid or at
query points, as in pixel-space diffusion\cite{hoogeboom2023simple} and
function-space generative models.\cite{kerrigan2024functional,lim2025score} 
This choice of latent or ambient representation influences how
much information passes from measured to unmeasured channels, and which scales a
sample can carry. On the cylinder, the low-dimensional
shedding state ties pressure to the measured velocities closely because all
channels carry similar structure, as in extended POD,\cite{boree2003extended} so
every method that models cross-channel structure recovers the never-observed
pressure. Isotropic turbulence, in contrast, is observed only in $U_x$ and $U_z$, at 1\%
sensing. No method recovers either unobserved channel well: the best errors,
0.710 on $U_y$ and 0.643 on $p$ (Latent FM, Table~\ref{tab:jhu}), are an order
of magnitude above the unobserved cylinder pressure error. Kinematics alone leaves
most of both channels undetermined on these subdomains. Given the true $U_x$
and $U_z$ everywhere, continuity fixes $U_y$ only up to its average along $y$,
which holds 78\% of the norm of $U_y$, so the best estimate from continuity
scores 0.79. Given the true velocity everywhere but no boundary pressure, the
pressure Poisson equation gives 0.78 on $p$ (SM, Sec.~S4).

Latent FM beats both bounds from 1\% sensing, so part of its information comes
from learned statistics rather than kinematics. It is the only method of
Table~\ref{tab:jhu} below 0.9 on both unobserved channels. Its advantage
comes from its representation. A decoder trained to reproduce complete 
four-channel fields can carry information from
measured to unmeasured channels, whereas Senseiver's latent array, which encodes
only the sensors and no global field context, shows no such advantage. 
Latent FM achieves lower errors on the unobserved channels than the 
kinematics-only reconstruction, despite using much sparser observations.
The improvement suggests that learned cross-channel statistics may provide 
information beyond what is obtained from the kinematic constraints alone.
This behavior is consistent with a latent decoder trained to represent 
complete multichannel flow fields, which may exploit correlations between 
measured and unmeasured variables.
This gain sits at the largest scales. Over the test half, Latent FM's $p$ 
correlates with the
truth at a median 0.88 for $k\le 7$, against 0.46 for FNO3D, 0.43 for SiT-point
and at most 0.38 for FNO3D-FM and Senseiver. On the frame of 
Fig.~\ref{fig:qual-jhu} its correlation at $k=8$--31 is 0.20 on $p$ and 0.03 on $U_y$.
Thus Latent FM exhibits advantages under severe under-determination.

\begin{figure}[tbp]
\includegraphics[width=\linewidth]{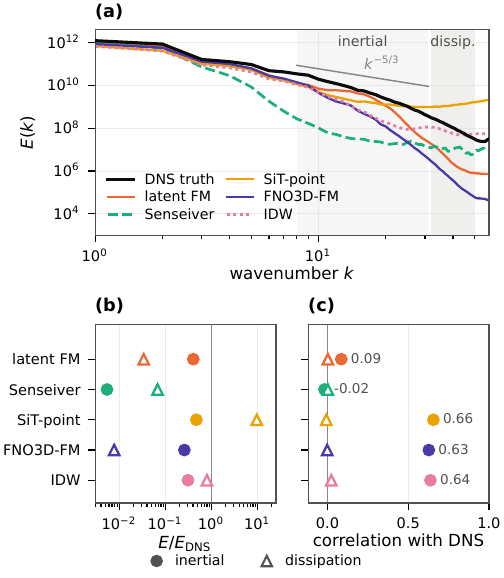}
\caption{\label{fig:spectra}Spectral content of single samples on the held-out
3D cube (frame of Fig.~\ref{fig:qual-jhu}, one sensor draw, sampler settings of
Table~\ref{tab:jhu}, Hann window).
(a) Energy spectra of the total velocity with the inertial ($k=8$--31) and
dissipation ($k=32$--50) bands shaded. (b) Band energy relative to the DNS and
(c) band-limited correlation with the DNS, pooled over the observed channels.
SiT-point members carry grid-scale speckle, which shows as excess
dissipation-band energy.}
\end{figure}

The representation also limits which scales a sample can infer from the sensors 
(Fig.~\ref{fig:spectra}). On the observed channels at $k=8$--31, the median
band correlation of Latent FM with the DNS over the test half is 0.04, against
0.65 for SiT-point and 0.63 for IDW. Latent FM 
receives the sensors as averages over $8^3$-point 
latent cells, so at 1\% sensing each cell holds about five sensors; the averaging 
removes structure that the sensors resolve between the cell scale ($k\approx8$) 
and the sensor spacing ($k\approx13$). Nevertheless, Latent FM keeps 0.49 of the DNS energy 
in this band, but the structure is randomly generated. On Kolmogorov flow, 
where the latent cells are finer than the sensor spacing, the band correlation 
of Latent FM at $k=16$--31 is 0.11.
In the dissipation band ($k=32$--50) Latent FM keeps 0.06 of the DNS energy, 
and FNO3D-FM and FNO3D, whose Fourier
layers stop above wavenumber eight, 0.01 each, so the cutoff is specific neither to
autoencoders nor to the objective. SiT-point's speckle puts 12 times the 
DNS energy in this band. No method's dissipation-band content correlates with the truth (at
most 0.02), because these wavelengths are shorter than the sensor spacing and 
the learned prior fixes at best their statistics. 
On Kolmogorov flow the latent-FM
stage 1 autoencoder alone keeps 0.95--0.99 of the small-scale energy, yet its stage 2
generated samples keep 0.30, so the loss arises in conditional generation rather than compression.
For the JHTDB cases, the 40-fold-compressed autoencoder alone keeps 0.98 of the
inertial-band energy of the true test-half fields, at a band correlation of
0.99, and reconstructs the four channels to relative $L_2$ errors of 0.031,
0.037, 0.033 and 0.025, so the decorrelation of the samples in this band does
not arise in compression. In the dissipation band it keeps 0.17 of the energy
(0.13--0.21 over the test half), against 0.06 for the samples. Compression 
therefore caps this band, similar to the deficit reported by Rafiq and
Nair,\cite{rafiq2026spectrally} and conditional generation removes about two
thirds of the remainder. Bulk metrics miss these losses: Latent FM leads the
aggregate $L_2$ of Table~\ref{tab:jhu}, and compression studies report preserved bulk
statistics and accuracy at 64- and 1000-fold compression.\cite{glaws2020deep,rozet2025lost} 
Representation covaries with backbone, conditioning and training budget, and between regimes
also with Reynolds number, dimension and geometry, so these mechanisms are
consistent with the results but are not controlled effects.

\subsection{Computational advantages reverse with resolution and query regime}\label{sec:cost}

Training takes 6.7--27.4 GPU-hours for every learned 3D method except the
early-stopped Set-DeepONet++ (1.0), but per-field inference time spans nearly 
four orders of magnitude, from
0.019\,s for Latent FM to about 150\,s for SiT-point (Table~\ref{tab:jhucost}).
Latent FM samples a compact latent state and decodes it once, whereas the slow
methods repeat many network passes through long guided chains (S3GM, CoNFiLD) or
sequential query chunks (SiT-point). All measurements are conducted with identical
GPUs so relative trends can be assessed. The scalability study uses synthetic 
inputs: each model reconstructs random fields on cubic lattices 
from $64^3$ to $1024^3$, with
the sensor count fixed at the 1\% budget of the $125^3$ cube, and
peak memory and wall time of one full-field reconstruction and one training
step are recorded. Memory and time depend on tensor shapes, not on values, so untrained
weights suffice for the grid models, which must be rebuilt at each resolution.

\begin{figure}[tbp]
\includegraphics[width=\linewidth]{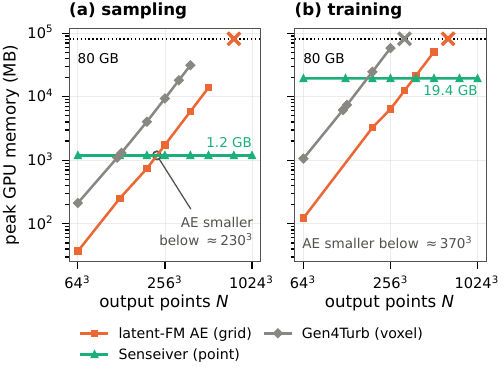}
\caption{\label{fig:scaling}Measured peak GPU memory against output resolution
(one H100, 80\,GB; log--log) for (a) full-field reconstruction and (b) one
training step. Crosses mark measured out-of-memory failures. Grid models run at
batch size 1; the latent-FM curve is its autoencoder alone, a lower bound on
sampling. The point model (Senseiver) evaluates fixed-size query chunks and stays
flat to $1024^3$.}
\end{figure}

For memory scalability, architectural differences become relevant.
Grid-native models compute on every point of the discretized field, so in 3D
their memory grows with the cube of the resolution
(Fig.~\ref{fig:scaling}, batch size 1). The convolutional autoencoder of latent
FM needs 0.25\,GB to encode and decode a $125^3$ field but runs out of memory at
$768^3$. In training, voxel diffusion (Gen4Turb) memory exceeds H100 limits (80~GB)
at $320^3$ and the Latent FM autoencoder at $640^3$.
Point-query models decouple working memory from the output resolution: they
train on a fixed subset of query points and evaluate a full field in fixed-size
chunks. The peak memory of Senseiver stays constant from $64^3$ to $1024^3$
(1.2\,GB for a full-field reconstruction and 22.4\,GB in training). Its run
time, however, grows with the number of queried points, from 0.125\,s at
$125^3$ to 67\,s at $1024^3$.
The two families therefore cross in memory and feasibility, not in run time. The
autoencoder uses less memory than Senseiver below about $230^3$ in
reconstruction, and wherever it fits its encode--decode is about 20 times faster
(6\,ms against 0.125\,s at $125^3$). Beyond the walls the full grid no longer
fits on one GPU, while the point model still runs at $1024^3$, and point models
also gain when only part of the field is needed.

No architecture is efficient in general: latent models pay for the latent
dimension and the decoded grid, grid-native models for the number of grid
points, point models for the number of queried points, and iterative samplers
for the number of network evaluations. The cheapest method depends on the target
resolution, the number of reconstructions and the fraction of each field that is
queried.

\section{Discussion and Model Selection}\label{sec:discussion}

No single method performs best across all three reconstruction regimes. 
Method selection should begin with architectural compatibility constraints 
(Sec.~\ref{sec:constraints}), including sensor geometry, query locations, 
output resolution, and memory requirements.
Among the admissible methods, the preferred approach depends on the information 
available from the sensors, the desired reconstruction quantities, and the 
computational budget (Sec.~\ref{sec:guidelines}).

\subsection{Hard architectural constraints}\label{sec:constraints}

Several of the tested methods contain hard limitations that prevent their use before
even assessing their accuracy. Latent FM, S3GM, Gen4Turb and the grid FNO need a 
Cartesian grid and thus are not tested on the body-fitted cylinder mesh 
(Table~\ref{tab:capability}). Set and point-cloud
models (Senseiver, SiT-point, MLP-RBF) accept any number of sensors
at arbitrary coordinates. Rasterizing models (Latent FM, FNO, Gen4Turb) move
sensors to grid nodes, and posterior-sampling models (S3GM, CoNFiLD) accept any
differentiable observation operator at a cost of seconds to minutes per field.
Full-grid decoders (Latent FM, S3GM, Gen4Turb, and FNO) return the full 
field at a fixed resolution, whereas point-query
models (Set-DeepONet, CoNFiLD, Senseiver, MLP-RBF, and SiT-point) 
evaluate arbitrary coordinates, such as probes, another mesh or a body
surface. Operator-based methods can make the solution resolution-invariant, in principle.
Full-grid models also hit measured memory walls between $320^3$ and
$768^3$ on one GPU, while point-query models still run at $1024^3$
(Sec.~\ref{sec:cost}).

The last filter is whether a deterministic estimate is enough or if generative
modeling should be used. A deterministic squared-error regressor
estimates the conditional mean, which suffices when the readings determine the
field and costs less than sampling. When several fields fit the readings, the
mean averages them. For a bimodal posterior it can fall between the modes, likewise
small-scale features seen in turbulence will be smoothed over.
\cite{blau2018perception,raonic2026gencfd} Applications that need
realistic members, spread or tail probabilities need a probabilistic
generative formulation.\cite{price2025probabilistic}

\subsection{Soft regime-dependent model-selection guidelines}\label{sec:guidelines}

Among the remaining methods, four questions may be used to guide the choice.
\begin{enumerate}
\item \emph{Will classical methods suffice?} The structure-function estimate
and the POD diagnostics need only training frames and the sensor layout
(Sec.~\ref{sec:density}, Table~\ref{tab:diagnostics}). If the predicted
nearest-neighbor error meets the accuracy requirement, IDW is sufficient. If a
few POD modes hold the variance and $H\Phi_r$ is well conditioned, gappy POD is
hard to beat, and LSE more so: they reach 0.057 and 0.043 from volume sensors of
the cylinder, against 0.59 and 0.34 (0.26 for ridge gappy POD) from 128 surface
taps of the same flow.
\item \emph{How dense is the sensing?} Over the Kolmogorov sweep, models that
keep a per-location record of the readings improve 5- to 16-fold (FNO,
SiT-point) and S3GM 21-fold between 1\% and 10\% sensing, although its sampler
diverges below 1\%. Models that pass the sensors through a fixed-size summary (Senseiver,
Latent FM, MLP-RBF) improve 2.5- to 3.6-fold and fall behind IDW at 10\%
sensing.
\item \emph{Must unobserved channels or fine scales be inferred?} The two needs
favor different representations. A latent decoder trained on complete
multichannel fields carries large-scale information to unmeasured channels,
but latent cells coarser than the sensor spacing discard scales that the
sensors resolve (Sec.~\ref{sec:identifiability}). No method infers scales
finer than the sensor spacing; a generative prior at best restores their energy.
\item \emph{How many outputs are queried?} Single-pass latent decoders are
fastest for repeated full-field reconstruction at moderate resolution. Point-query
models are preferable when only part of the field is needed, and iterative
samplers pay per network evaluation in any representation (Sec.~\ref{sec:cost}).
\end{enumerate}

\subsection{Limitations}\label{sec:limitations}

Limitations exist in the present work that simultaneously are opportunity for
later investigations. All three regimes are simulations of canonical flows on two 
Cartesian grids and one body-fitted O-grid, and the suite contains no experimental data, so 
neither measurement realism nor geometric generality can be claimed here. Sensor 
noise is synthetic and tested on one 2D regime (SM, Sec.~S2), and the surface study uses
one body. Every learned method trains per regime and reconstructs single
snapshots, whereas variational assimilation of observation histories has
recovered all scales of channel turbulence.\cite{zaki2021limited}

\section{Conclusion}\label{sec:conclusion}

\benchname{} evaluates state of the art sparse-sensing data-driven techniques under a common protocol. 
Three regimes are tested, from 2D laminar to 3D turbulent flow, ten learned 
methods and five classical baselines, including leakage-controlled splits, proper  probabilistic scores, member-level physics metrics, and practical computational cost. 
While no method leads in every regime, trends are revealed between methods that guide method selection. 
Classical baselines are hard to beat when the sensors are dense relative
to the field's decorrelation length or observe a compressible POD basis, and
simple diagnostics predict both cases. Mean-field error alone misranks
generative methods, most raw ensembles are under-dispersed at error-selected
sampler settings, and a scalar recalibration transfers only when the
miscalibration is stationary. On the 3D subdomains, kinematics leaves most of
the unobserved channels undetermined, and representation choices trade
cross-channel coupling, fine-scale content and scaling. The datasets, splits,
sensor draws and evaluation code will be released for extension.

\section*{Supplementary Material}
See the supplementary material for the seed-repeat statistics (Sec.~S1), the
sensor-noise and occlusion tests (Sec.~S2), the Gen4Turb channel asymmetry and
the kinematic bounds on the unobserved 3D channels (Sec.~S4), the leakage test
on training-trajectory frames (Sec.~S6), the comparison of LSE with gappy POD
and the conditional recalibration (Sec.~S7), the parameter-count-matched CoNFiLD
runs, the per-Reynolds-number cylinder errors (Table~S2), further CoNFiLD,
SiT-point and S3GM tests on the 3D cube (Table~S3), optimizer steps and time per
step of the 3D methods (Table~S4), the CRPS-optimal sampler settings
(Table~S5), and the 3D sensor-density sweep (Fig.~S1).

\begin{acknowledgments}
This material is based upon work supported by the U.S. National Science
Foundation under Award No.~CBET-143625.
N.J.T. is partially supported by GE Vernova under the GE Vernova X MIT fellowship.
J.C. is partially supported by the Jack Kent Cooke Foundation.
This research was performed using computational resources sponsored by the U.S. Department of Energy's Office of Critical Minerals and Energy Innovation and located at the National Laboratory of the Rockies.
Any opinions, findings, and conclusions or
recommendations expressed in this material are those of the authors and do not
necessarily reflect the views of the National Science Foundation.
\end{acknowledgments}

\section*{Author Declarations}
\subsection*{Conflict of Interest}
The authors have no conflicts to disclose.

\subsection*{Author Contributions}
Nicolas Tricard: Conceptualization, Formal analysis, Data curation, Investigation, Methodology, Software, Validation, Visualization, Writing – original draft. Linzheng Wang: Conceptualization, Methodology, Software, Writing – review \& editing. Jason Chen: Software, Writing – review \& editing. Sili Deng: Conceptualization, Funding acquisition, Project administration, Supervision, Writing – review \& editing.

\section*{Data Availability Statement}
The cylinder-wake dataset generated in this study, the dataset splits, the
evaluation sensor draws and the benchmark code are available from the
corresponding author upon reasonable request and will be released in a public
repository upon publication. The two-dimensional Kolmogorov flow data are from
Ref.~\onlinecite{shu2023physics}, and the three-dimensional isotropic turbulence
data were obtained from the Johns Hopkins Turbulence Databases.\cite{li2008public}

\bibliography{references,refs_extra}

@article{fukami2021global,
  title   = {Global field reconstruction from sparse sensors with {V}oronoi tessellation-assisted deep learning},
  author  = {Fukami, Kai and Maulik, Romit and Ramachandra, Nesar and Fukagata, Koji and Taira, Kunihiko},
  journal = {Nat. Mach. Intell.},
  volume  = {3},
  number  = {11},
  pages   = {945--951},
  year    = {2021}
}

@article{santos2023senseiver,
  title   = {Development of the {S}enseiver for efficient field reconstruction from sparse observations},
  author  = {Santos, Javier E and Fox, Zachary R and Mohan, Arvind and O'Malley, Daniel and Viswanathan, Hari and Lubbers, Nicholas},
  journal = {Nat. Mach. Intell.},
  volume  = {5},
  number  = {11},
  pages   = {1317--1325},
  year    = {2023}
}

@misc{dang2025flronet,
  author        = {Dang, Hiep Vo and Choi, Joseph B. and Nguyen, Phong C. H.},
  title         = {{FLRONet}: Deep operator learning for high-fidelity fluid flow field reconstruction from sparse sensor measurements},
  year          = {2025},
  eprint        = {2412.08009v3},
  archiveprefix = {arXiv},
  primaryclass  = {physics.flu-dyn}
}

@article{erichson2020shallow,
  title   = {Shallow neural networks for fluid flow reconstruction with limited sensors},
  author  = {Erichson, N Benjamin and Mathelin, Lionel and Yao, Zhewei and Brunton, Steven L and Mahoney, Michael W and Kutz, J Nathan},
  journal = {Proc. R. Soc. A},
  volume  = {476},
  number  = {2238},
  pages   = {20200097},
  year    = {2020}
}

@article{guastoni2021convolutional,
  title   = {Convolutional-network models to predict wall-bounded turbulence from wall quantities},
  author  = {Guastoni, Luca and G{\"u}emes, Alejandro and Ianiro, Andrea and Discetti, Stefano and Schlatter, Philipp and Azizpour, Hossein and Vinuesa, Ricardo},
  journal = {J. Fluid Mech.},
  volume  = {928},
  pages   = {A27},
  year    = {2021}
}

@article{cuellar2024three,
  title   = {Three-dimensional generative adversarial networks for turbulent flow estimation from wall measurements},
  author  = {Cu{\'e}llar, Antonio and G{\"u}emes, Alejandro and Ianiro, Andrea and Flores, \'{O}scar and Vinuesa, Ricardo and Discetti, Stefano},
  journal = {J. Fluid Mech.},
  volume  = {991},
  pages   = {A1},
  year    = {2024}
}

@article{callaham2019robust,
  title   = {Robust flow reconstruction from limited measurements via sparse representation},
  author  = {Callaham, Jared L and Maeda, Kazuki and Brunton, Steven L},
  journal = {Phys. Rev. Fluids},
  volume  = {4},
  number  = {10},
  pages   = {103907},
  year    = {2019}
}

@article{li2024s3gm,
  title   = {Learning spatiotemporal dynamics with a pretrained generative model},
  author  = {Li, Zeyu and Han, Wang and Zhang, Yue and Fu, Qingfei and Li, Jingxuan and Qin, Lizi and Dong, Ruoyu and Sun, Hao and Deng, Yue and Yang, Lijun},
  journal = {Nat. Mach. Intell.},
  volume  = {6},
  number  = {12},
  pages   = {1566--1579},
  year    = {2024}
}

@article{du2024confild,
  title   = {Conditional neural field latent diffusion model for generating spatiotemporal turbulence},
  author  = {Du, Pan and Parikh, Meet Hemant and Fan, Xiantao and Liu, Xin-Yang and Wang, Jian-Xun},
  journal = {Nat. Commun.},
  volume  = {15},
  pages   = {10416},
  year    = {2024}
}

@article{oommen2026turbulent,
  title   = {Learning turbulent flows with generative models for super-resolution and sparse flow reconstruction},
  author  = {Oommen, Vivek and Khodakarami, Siavash and Bora, Aniruddha and Wang, Zhicheng and Karniadakis, George Em},
  journal = {Nat. Commun.},
  volume  = {17},
  pages   = {3707},
  year    = {2026}
}

@article{amoros2026guiding,
  title   = {Guiding diffusion models to reconstruct flow fields from sparse data},
  author  = {Amor{\'o}s-Trepat, Marc and Medrano-Navarro, Luis and Liu, Qiang and Guastoni, Luca and Thuerey, Nils},
  journal = {Phys. Fluids},
  volume  = {38},
  number  = {1},
  pages   = {015112},
  year    = {2026}
}

@article{shu2023physics,
  title   = {A physics-informed diffusion model for high-fidelity flow field reconstruction},
  author  = {Shu, Dule and Li, Zijie and Farimani, Amir Barati},
  journal = {J. Comput. Phys.},
  volume  = {478},
  pages   = {111972},
  year    = {2023}
}

@inproceedings{huang2024diffusionpde,
  title     = {{DiffusionPDE}: Generative {PDE}-solving under partial observation},
  author    = {Huang, Jiahe and Yang, Guandao and Wang, Zichen and Park, Jeong Joon},
  booktitle = {Advances in Neural Information Processing Systems},
  volume    = {37},
  year      = {2024},
  pages     = {130291--130323},
  doi       = {10.52202/079017-4140}
}

@inproceedings{kerrigan2024functional,
  title     = {Functional flow matching},
  author    = {Kerrigan, Gavin and Migliorini, Giosue and Smyth, Padhraic},
  booktitle = {Proceedings of the 27th International Conference on Artificial Intelligence and Statistics},
  pages     = {3934--3942},
  year      = {2024},
  series    = {Proceedings of Machine Learning Research},
  volume    = {238}
}

@article{lim2025score,
  title   = {Score-based diffusion models in function space},
  author  = {Lim, Jae Hyun and Kovachki, Nikola B and Baptista, Ricardo and Beckham, Christopher and Azizzadenesheli, Kamyar and Kossaifi, Jean and Voleti, Vikram and Song, Jiaming and Kreis, Karsten and Kautz, Jan and Pal, Christopher and Vahdat, Arash and Anandkumar, Anima},
  journal = {J. Mach. Learn. Res.},
  volume  = {26},
  number  = {158},
  pages   = {1--62},
  year    = {2025}
}

@inproceedings{lipman2023flow,
  title     = {Flow matching for generative modeling},
  author    = {Lipman, Yaron and Chen, Ricky T. Q. and Ben-Hamu, Heli and Nickel, Maximilian and Le, Matt},
  booktitle = {International Conference on Learning Representations},
  year      = {2023}
}

@inproceedings{liu2023flow,
  title     = {Flow straight and fast: Learning to generate and transfer data with rectified flow},
  author    = {Liu, Xingchao and Gong, Chengyue and Liu, Qiang},
  booktitle = {International Conference on Learning Representations},
  year      = {2023}
}

@misc{rafiq2026spectrally,
  title   = {Spectrally regularized latent flow matching for turbulence generation},
  author  = {Rafiq, Khalid and Nair, Aditya G},
  year    = {2026},
  eprint  = {2606.11691},
  archivePrefix = {arXiv}
}

@article{glaws2020deep,
  title   = {Deep learning for in situ data compression of large turbulent flow simulations},
  author  = {Glaws, Andrew and King, Ryan and Sprague, Michael},
  journal = {Phys. Rev. Fluids},
  volume  = {5},
  number  = {11},
  pages   = {114602},
  year    = {2020}
}

@inproceedings{li2021fourier,
  title     = {Fourier neural operator for parametric partial differential equations},
  author    = {Li, Zongyi and Kovachki, Nikola B and Azizzadenesheli, Kamyar and Liu, Burigede and Bhattacharya, Kaushik and Stuart, Andrew and Anandkumar, Anima},
  booktitle = {International Conference on Learning Representations},
  year      = {2021}
}

@article{kovachki2023neural,
  title   = {Neural operator: Learning maps between function spaces with applications to {PDE}s},
  author  = {Kovachki, Nikola B and Li, Zongyi and Liu, Burigede and Azizzadenesheli, Kamyar and Bhattacharya, Kaushik and Stuart, Andrew and Anandkumar, Anima},
  journal = {J. Mach. Learn. Res.},
  volume  = {24},
  number  = {89},
  pages   = {1--97},
  year    = {2023}
}

@inproceedings{jaegle2021perceiver,
  title     = {Perceiver: General perception with iterative attention},
  author    = {Jaegle, Andrew and Gimeno, Felix and Brock, Andy and Vinyals, Oriol and Zisserman, Andrew and Carreira, Joao},
  booktitle = {Proceedings of the 38th International Conference on Machine Learning},
  pages     = {4651--4664},
  year      = {2021},
  series    = {Proceedings of Machine Learning Research},
  volume    = {139}
}

@inproceedings{ma2024sit,
  title     = {{SiT}: Exploring flow and diffusion-based generative models with scalable interpolant transformers},
  author    = {Ma, Nanye and Goldstein, Mark and Albergo, Michael S and Boffi, Nicholas M and Vanden-Eijnden, Eric and Xie, Saining},
  booktitle = {Computer Vision -- {ECCV} 2024},
  year      = {2024},
  series    = {Lecture Notes in Computer Science},
  volume    = {15135},
  pages     = {23--40},
  publisher = {Springer},
  address   = {Cham},
  doi       = {10.1007/978-3-031-72980-5_2}
}

@inproceedings{rombach2022high,
  title     = {High-resolution image synthesis with latent diffusion models},
  author    = {Rombach, Robin and Blattmann, Andreas and Lorenz, Dominik and Esser, Patrick and Ommer, Bj{\"o}rn},
  booktitle = {Proceedings of the {IEEE/CVF} Conference on Computer Vision and Pattern Recognition ({CVPR})},
  pages     = {10674--10685},
  year      = {2022},
  doi       = {10.1109/CVPR52688.2022.01042}
}

@article{vinuesa2023transformative,
  title   = {The transformative potential of machine learning for experiments in fluid mechanics},
  author  = {Vinuesa, Ricardo and Brunton, Steven L and McKeon, Beverley J},
  journal = {Nat. Rev. Phys.},
  volume  = {5},
  number  = {9},
  pages   = {536--545},
  year    = {2023}
}

@article{mousavi2025low,
  title   = {Low-order flow reconstruction and uncertainty quantification in disturbed aerodynamics using sparse pressure measurements},
  author  = {Mousavi, Hanieh and Eldredge, Jeff D},
  journal = {J. Fluid Mech.},
  volume  = {1013},
  pages   = {A41},
  year    = {2025}
}

@article{raissi2020hidden,
  title   = {Hidden fluid mechanics: Learning velocity and pressure fields from flow visualizations},
  author  = {Raissi, Maziar and Yazdani, Alireza and Karniadakis, George Em},
  journal = {Science},
  volume  = {367},
  number  = {6481},
  pages   = {1026--1030},
  year    = {2020}
}

@article{li2008public,
  title   = {A public turbulence database cluster and applications to study {L}agrangian evolution of velocity increments in turbulence},
  author  = {Li, Yi and Perlman, Eric and Wan, Minping and Yang, Yunke and Meneveau, Charles and Burns, Randal and Chen, Shiyi and Szalay, Alexander and Eyink, Gregory},
  journal = {J. Turbul.},
  volume  = {9},
  pages   = {N31},
  year    = {2008}
}

@inproceedings{ohana2024well,
  title     = {The {W}ell: a large-scale collection of diverse physics simulations for machine learning},
  author    = {Ohana, Ruben and McCabe, Michael and Meyer, Lucas and Morel, Rudy and Agocs, Fruzsina J. and Beneitez, Miguel and Berger, Marsha and Burkhart, Blakesley and Burns, Keaton and Dalziel, Stuart B. and Fielding, Drummond B. and Fortunato, Daniel and Goldberg, Jared A. and Hirashima, Keiya and Jiang, Yan-Fei and Kerswell, Rich R. and Maddu, Suryanarayana and Miller, Jonah and Mukhopadhyay, Payel and Nixon, Stefan S. and Shen, Jeff and Watteaux, Romain and R{\'e}galdo-Saint Blancard, Bruno and Rozet, Fran{\c{c}}ois and Parker, Liam H. and Cranmer, Miles and Ho, Shirley},
  booktitle = {Advances in Neural Information Processing Systems},
  volume    = {37},
  year      = {2024},
  pages     = {44989--45037},
  doi       = {10.52202/079017-1430}
}

@article{gneiting2007strictly,
  title   = {Strictly proper scoring rules, prediction, and estimation},
  author  = {Gneiting, Tilmann and Raftery, Adrian E},
  journal = {J. Am. Stat. Assoc.},
  volume  = {102},
  number  = {477},
  pages   = {359--378},
  year    = {2007}
}

@inproceedings{chung2023diffusion,
  title     = {Diffusion posterior sampling for general noisy inverse problems},
  author    = {Chung, Hyungjin and Kim, Jeongsol and {McCann}, Michael T. and Klasky, Marc L and Ye, Jong Chul},
  booktitle = {International Conference on Learning Representations},
  year      = {2023}
}

@article{rybchuk2023ensemble,
  title   = {Ensemble flow reconstruction in the atmospheric boundary layer from spatially limited measurements through latent diffusion models},
  author  = {Rybchuk, Alex and Hassanaly, Malik and Hamilton, Nicholas and Doubrawa, Paula and Fulton, Mitchell J and Mart{\'i}nez-Tossas, Luis A},
  journal = {Phys. Fluids},
  volume  = {35},
  number  = {12},
  pages   = {126604},
  year    = {2023}
}

@article{manshausen2025generative,
  title   = {Generative data assimilation of sparse weather station observations at kilometer scales},
  author  = {Manshausen, Peter and Cohen, Yair and Harrington, Peter and Pathak, Jaideep and Pritchard, Mike and Garg, Piyush and Mardani, Morteza and Kashinath, Karthik and Byrne, Simon and Brenowitz, Noah},
  journal = {J. Adv. Model. Earth Syst.},
  volume  = {17},
  number  = {10},
  pages   = {e2024MS004505},
  year    = {2025}
}

@article{raonic2026gencfd,
  title   = {Generative {AI} for efficient statistical computation of fluids},
  author  = {Raoni{\'c}, Bogdan and Molinaro, Roberto and Lanthaler, Samuel and Rohner, Tobias and Armegioiu, Victor and Simonis, Stephan and Grund, Dana and Ramic, Yannick and Wan, Zhong Yi and Sha, Fei and Mishra, Siddhartha and Zepeda-N{\'u}{\~n}ez, Leonardo},
  journal = {Nat. Commun.},
  year    = {2026},
  doi     = {10.1038/s41467-026-76390-x},
  volume  = {17},
  pages   = {9846}
}

@misc{yao2026posteriorbenchpointestimatesposterior,
  title         = {{PosteriorBench}: From Point Estimates to Posterior Matching in Evaluating Generative Inverse Solvers},
  author        = {Jiachen Yao and Zi-Siang Hsu and Xi Deng and Aditi Gupta and Xin Ju and Sally M Benson and Gege Wen and Anima Anandkumar},
  year          = {2026},
  eprint        = {2609.20794},
  archiveprefix = {arXiv},
  primaryclass  = {cs.LG},
  url           = {https://arxiv.org/abs/2609.20794}
}

@inproceedings{karras2022elucidatingdesignspacediffusionbased,
  title     = {Elucidating the design space of diffusion-based generative models},
  author    = {Karras, Tero and Aittala, Miika and Aila, Timo and Laine, Samuli},
  year      = {2022},
  booktitle = {Advances in Neural Information Processing Systems},
  volume    = {35},
  pages     = {26565--26577},
  doi       = {10.52202/068431-1926}
}

@article{Wang2026Learning,
  title = {Learning multi-physical system on a unified manifold by collaboratively fused features},
  volume = {170},
  ISSN = {0952-1976},
  url = {http://dx.doi.org/10.1016/j.engappai.2026.114178},
  DOI = {10.1016/j.engappai.2026.114178},
  journal = {Eng. Appl. Artif. Intell.},
  publisher = {Elsevier BV},
  author = {Wang,  Linzheng and Chen,  Zituo and Li,  Yaojun and Deng,  Ruiqu and Zhang,  Ruizhi and Luo,  Yonghao and Lu,  Lu and Deng,  Sili},
  year = {2026},
  month = Apr,
  pages = {114178}
}

@article{williamson1996vortex,
  author  = {Williamson, C. H. K.},
  title   = {Vortex dynamics in the cylinder wake},
  journal = {Annu. Rev. Fluid Mech.},
  volume  = {28},
  pages   = {477--539},
  year    = {1996}
}

@article{norberg2003fluctuating,
  author  = {Norberg, C.},
  title   = {Fluctuating lift on a circular cylinder: review and new measurements},
  journal = {J. Fluids Struct.},
  volume  = {17},
  number  = {1},
  pages   = {57--96},
  year    = {2003}
}

@article{everson1995karhunen,
  author  = {Everson, R. and Sirovich, L.},
  title   = {{Karhunen--Lo{\`e}ve} procedure for gappy data},
  journal = {J. Opt. Soc. Am. A},
  volume  = {12},
  number  = {8},
  pages   = {1657--1664},
  year    = {1995}
}

@article{lu2021learning,
  author  = {Lu, L. and Jin, P. and Pang, G. and Zhang, Z. and Karniadakis, G. E.},
  title   = {Learning nonlinear operators via {DeepONet} based on the universal approximation theorem of operators},
  journal = {Nat. Mach. Intell.},
  volume  = {3},
  number  = {3},
  pages   = {218--229},
  year    = {2021}
}

@misc{wang2026dmfgen,
  author        = {Wang, Linzheng and Chen, Jason and Tricard, Nicolas and Chen, Zituo and Guo, Xingsen and Deng, Sili},
  title         = {Sparse reconstruction by direct measurement-to-field generation in function space},
  year          = {2026},
  eprint        = {2610.34709},
  archiveprefix = {arXiv},
  primaryclass  = {cs.LG},
  doi           = {10.48550/arXiv.2610.34709}
}

@inproceedings{takamoto2022pdebench,
  author    = {Takamoto, Makoto and Praditia, Timothy and Leiteritz, Raphael and MacKinlay, Dan and Alesiani, Francesco and Pfl{\"u}ger, Dirk and Niepert, Mathias},
  title     = {{PDEBench}: An extensive benchmark for scientific machine learning},
  booktitle = {Advances in Neural Information Processing Systems},
  volume    = {35},
  pages     = {1596--1611},
  year      = {2022},
  doi       = {10.52202/068431-0117}
}

@misc{luo2023cfdbench,
  author  = {Luo, Yining and Chen, Yingfa and Zhang, Zhen},
  title   = {{CFDBench}: A large-scale benchmark for machine learning methods in fluid dynamics},
  year    = {2023},
  eprint  = {2310.05963},
  archivePrefix = {arXiv}
}

@inproceedings{bonnet2022airfrans,
  author    = {Bonnet, Florent and Mazari, Jocelyn Ahmed and Cinnella, Paola and Gallinari, Patrick},
  title     = {{AirfRANS}: High fidelity computational fluid dynamics dataset for approximating {Reynolds}-averaged {Navier}--{Stokes} solutions},
  booktitle = {Advances in Neural Information Processing Systems},
  volume    = {35},
  pages     = {23463--23478},
  year      = {2022},
  doi       = {10.52202/068431-1705}
}

@article{li2023geofno,
  author  = {Li, Zongyi and Huang, Daniel Zhengyu and Liu, Burigede and Anandkumar, Anima},
  title   = {{Fourier} neural operator with learned deformations for {PDEs} on general geometries},
  journal = {J. Mach. Learn. Res.},
  volume  = {24},
  number  = {388},
  pages   = {1--26},
  year    = {2023}
}

@inproceedings{blau2018perception,
  title     = {The Perception-Distortion Tradeoff},
  author    = {Blau, Yochai and Michaeli, Tomer},
  booktitle = {Proceedings of the IEEE Conference on Computer Vision and Pattern Recognition (CVPR)},
  pages     = {6228--6237},
  year      = {2018},
  doi       = {10.1109/CVPR.2018.00652}
}

@article{price2025probabilistic,
  title   = {Probabilistic weather forecasting with machine learning},
  author  = {Price, Ilan and Sanchez-Gonzalez, Alvaro and Alet, Ferran and Andersson, Tom R. and El-Kadi, Andrew and Masters, Dominic and Ewalds, Timo and Stott, Jacklynn and Mohamed, Shakir and Battaglia, Peter and Lam, Remi and Willson, Matthew},
  journal = {Nature},
  volume  = {637},
  number  = {8044},
  pages   = {84--90},
  year    = {2025}
}

@article{ravuri2021skilful,
  title   = {Skilful precipitation nowcasting using deep generative models of radar},
  author  = {Ravuri, S. and Lenc, K. and Willson, M. and Kangin, D. and Lam, R. and Mirowski, P. and Fitzsimons, M. and Athanassiadou, M. and Kashem, S. and Madge, S. and Prudden, R. and Mandhane, A. and Clark, A. and Brock, A. and Simonyan, K. and Hadsell, R. and Robinson, N. and Clancy, E. and Arribas, A. and Mohamed, S.},
  journal = {Nature},
  volume  = {597},
  number  = {7878},
  pages   = {672--677},
  year    = {2021}
}

@article{fortin2014why,
  title   = {Why should ensemble spread match the {RMSE} of the ensemble mean?},
  author  = {Fortin, V. and Abaza, M. and Anctil, F. and Turcotte, R.},
  journal = {J. Hydrometeorol.},
  volume  = {15},
  number  = {4},
  pages   = {1708--1713},
  year    = {2014}
}

@article{matheron1963principles,
  author  = {Matheron, G.},
  title   = {Principles of geostatistics},
  journal = {Econ. Geol.},
  volume  = {58},
  number  = {8},
  pages   = {1246--1266},
  year    = {1963}
}

@article{zimmerman1999experimental,
  author  = {Zimmerman, D. and Pavlik, C. and Ruggles, A. and Armstrong, M. P.},
  title   = {An experimental comparison of ordinary and universal kriging and inverse distance weighting},
  journal = {Math. Geol.},
  volume  = {31},
  number  = {4},
  pages   = {375--390},
  year    = {1999}
}

@article{kravchenko2003influence,
  author  = {Kravchenko, A. N.},
  title   = {Influence of spatial structure on accuracy of interpolation methods},
  journal = {Soil Sci. Soc. Am. J.},
  volume  = {67},
  number  = {5},
  pages   = {1564--1571},
  year    = {2003}
}

@article{zhang2015spatial,
  author  = {Zhang, H. and Lu, L. and Liu, Y. and Liu, W.},
  title   = {Spatial sampling strategies for the effect of interpolation accuracy},
  journal = {ISPRS Int. J. Geo-Inf.},
  volume  = {4},
  number  = {4},
  pages   = {2742--2768},
  year    = {2015}
}

@article{willcox2006unsteady,
  author  = {Willcox, K.},
  title   = {Unsteady flow sensing and estimation via the gappy proper orthogonal decomposition},
  journal = {Comput. Fluids},
  volume  = {35},
  number  = {2},
  pages   = {208--226},
  year    = {2006}
}

@article{peherstorfer2020stability,
  author  = {Peherstorfer, Benjamin and Drma{\v{c}}, Zlatko and Gugercin, Serkan},
  title   = {Stability of discrete empirical interpolation and gappy proper orthogonal decomposition with randomized and deterministic sampling points},
  journal = {SIAM J. Sci. Comput.},
  volume  = {42},
  number  = {5},
  pages   = {A2837--A2864},
  year    = {2020}
}

@article{manohar2018data,
  author  = {Manohar, Krithika and Brunton, Bingni W. and Kutz, J. Nathan and Brunton, Steven L.},
  title   = {Data-driven sparse sensor placement for reconstruction: Demonstrating the benefits of exploiting known patterns},
  journal = {IEEE Control Syst. Mag.},
  volume  = {38},
  number  = {3},
  pages   = {63--86},
  year    = {2018}
}

@article{zhang2025operator,
  title   = {Operator learning for reconstructing flow fields from sparse measurements: an energy transformer approach},
  author  = {Zhang, Qian and Krotov, Dmitry and Karniadakis, George Em},
  journal = {J. Comput. Phys.},
  volume  = {538},
  pages   = {114148},
  year    = {2025}
}

@article{boree2003extended,
  title   = {Extended proper orthogonal decomposition: a tool to analyse correlated events in turbulent flows},
  author  = {Bor{\'e}e, J.},
  journal = {Exp. Fluids},
  volume  = {35},
  number  = {2},
  pages   = {188--192},
  year    = {2003}
}

@article{zaki2021limited,
  title   = {From limited observations to the state of turbulence: Fundamental difficulties of flow reconstruction},
  author  = {Zaki, Tamer A. and Wang, Mengze},
  journal = {Phys. Rev. Fluids},
  volume  = {6},
  number  = {10},
  pages   = {100501},
  year    = {2021}
}

@inproceedings{rozet2025lost,
  title     = {Lost in latent space: An empirical study of latent diffusion models for physics emulation},
  author    = {Rozet, Fran{\c{c}}ois and Ohana, Ruben and McCabe, Michael and Louppe, Gilles and Lanusse, Fran{\c{c}}ois and Ho, Shirley},
  booktitle = {Advances in Neural Information Processing Systems},
  year      = {2025},
  volume    = {38},
  pages     = {149258--149303},
  doi       = {10.52202/085713-4492}
}

@inproceedings{hoogeboom2023simple,
  title     = {{simple diffusion}: End-to-end diffusion for high resolution images},
  author    = {Hoogeboom, Emiel and Heek, Jonathan and Salimans, Tim},
  booktitle = {Proceedings of the 40th International Conference on Machine Learning},
  pages     = {13213--13232},
  year      = {2023},
  series    = {Proceedings of Machine Learning Research},
  volume    = {202}
}

@inproceedings{kuleshov2018accurate,
  title     = {Accurate uncertainties for deep learning using calibrated regression},
  author    = {Kuleshov, Volodymyr and Fenner, Nathan and Ermon, Stefano},
  booktitle = {Proceedings of the 35th International Conference on Machine Learning},
  series    = {Proceedings of Machine Learning Research},
  volume    = {80},
  pages     = {2796--2804},
  year      = {2018}
}

@article{anderson2009spatially,
  title   = {Spatially and temporally varying adaptive covariance inflation for ensemble filters},
  author  = {Anderson, Jeffrey L},
  journal = {Tellus A},
  volume  = {61},
  number  = {1},
  pages   = {72--83},
  year    = {2009}
}

@article{li2023multiscale,
  author  = {Li, Tianyi and Buzzicotti, Michele and Biferale, Luca and Bonaccorso, Fabio and Chen, Shiyi and Wan, Minping},
  title   = {Multi-scale reconstruction of turbulent rotating flows with proper orthogonal decomposition and generative adversarial networks},
  journal = {J. Fluid Mech.},
  volume  = {971},
  pages   = {A3},
  year    = {2023}
}

@inproceedings{sitzmann2020siren,
  author    = {Sitzmann, Vincent and Martel, Julien and Bergman, Alexander and Lindell, David and Wetzstein, Gordon},
  booktitle = {Advances in Neural Information Processing Systems},
  pages     = {7462--7473},
  title     = {Implicit Neural Representations with Periodic Activation Functions},
  volume    = {33},
  year      = {2020}
}

@misc{zhou2026physicalfieldreconstructionsparseobservations,
  title         = {Physical-Field Reconstruction from Sparse Observations: When Are Diffusion Models Preferable to Deterministic Regression?},
  author        = {Hao Zhou and Rui Zhang and Qi Wang and Hao Sun},
  year          = {2026},
  eprint        = {2609.06135},
  archiveprefix = {arXiv},
  primaryclass  = {physics.comp-ph},
  url           = {https://arxiv.org/abs/2609.06135}
}

@article{STAHL2025114037,
  title   = {Sparse flow reconstruction methods to reduce the costs of analyzing large unsteady datasets},
  journal = {J. Comput. Phys.},
  volume  = {534},
  pages   = {114037},
  year    = {2025},
  issn    = {0021-9991},
  doi     = {10.1016/j.jcp.2025.114037},
  author  = {Spencer L. Stahl and Stuart I. Benton}
}

@inproceedings{shepard1968idw,
  author    = {Shepard, Donald},
  title     = {A two-dimensional interpolation function for irregularly-spaced data},
  year      = {1968},
  isbn      = {9781450374866},
  publisher = {Association for Computing Machinery},
  address   = {New York, NY, USA},
  url       = {https://doi.org/10.1145/800186.810616},
  doi       = {10.1145/800186.810616},
  booktitle = {Proceedings of the 1968 23rd ACM National Conference},
  pages     = {517--524},
  numpages  = {8},
  series    = {ACM '68}
}

@article{brunton2020machine,
  title   = {Machine learning for fluid mechanics},
  author  = {Brunton, Steven L. and Noack, Bernd R. and Koumoutsakos, Petros},
  journal = {Annu. Rev. Fluid Mech.},
  volume  = {52},
  pages   = {477--508},
  year    = {2020},
  doi     = {10.1146/annurev-fluid-010719-060214}
}

@article{fukami2023super,
  title   = {Super-resolution analysis via machine learning: a survey for fluid flows},
  author  = {Fukami, Kai and Fukagata, Koji and Taira, Kunihiko},
  journal = {Theor. Comput. Fluid Dyn.},
  volume  = {37},
  number  = {4},
  pages   = {421--444},
  year    = {2023},
  doi     = {10.1007/s00162-023-00663-0}
}

@article{comtebellot1976hot,
  title   = {Hot-wire anemometry},
  author  = {Comte-Bellot, Genevi{\`e}ve},
  journal = {Annu. Rev. Fluid Mech.},
  volume  = {8},
  pages   = {209--231},
  year    = {1976},
  doi     = {10.1146/annurev.fl.08.010176.001233}
}

@techreport{whitmore1995rtfads,
  title       = {In-flight demonstration of a real-time flush airdata sensing ({RT-FADS}) system},
  author      = {Whitmore, Stephen A. and Davis, Roy J. and Fife, John Michael},
  institution = {NASA Dryden Flight Research Center},
  number      = {NASA-TM-104314},
  year        = {1995}
}

@article{vanoudheusden2013piv,
  title   = {{PIV}-based pressure measurement},
  author  = {van Oudheusden, B. W.},
  journal = {Meas. Sci. Technol.},
  volume  = {24},
  number  = {3},
  pages   = {032001},
  year    = {2013},
  doi     = {10.1088/0957-0233/24/3/032001}
}

@article{stuart2010inverse,
  title   = {Inverse problems: A {B}ayesian perspective},
  author  = {Stuart, Andrew M.},
  journal = {Acta Numer.},
  volume  = {19},
  pages   = {451--559},
  year    = {2010},
  doi     = {10.1017/S0962492910000061}
}

@article{berkooz1993proper,
  title   = {The proper orthogonal decomposition in the analysis of turbulent flows},
  author  = {Berkooz, Gal and Holmes, Philip and Lumley, John L.},
  journal = {Annu. Rev. Fluid Mech.},
  volume  = {25},
  pages   = {539--575},
  year    = {1993},
  doi     = {10.1146/annurev.fl.25.010193.002543}
}

@article{taira2017modal,
  title   = {Modal analysis of fluid flows: An overview},
  author  = {Taira, Kunihiko and Brunton, Steven L. and Dawson, Scott T. M. and Rowley, Clarence W. and Colonius, Tim and McKeon, Beverley J. and Schmidt, Oliver T. and Gordeyev, Stanislav and Theofilis, Vassilios and Ukeiley, Lawrence S.},
  journal = {AIAA J.},
  volume  = {55},
  number  = {12},
  pages   = {4013--4041},
  year    = {2017},
  doi     = {10.2514/1.J056060}
}

@misc{huang2026fluidsplat,
  title   = {{FLUIDSPLAT}: Reconstructing physical fields from sparse sensors via {G}aussian primitives},
  author  = {Huang, Huaxi and Li, Meng and Gao, Zhengqing and Zhou, Xi and Huang, Xiaoshui and Sun, Xiao},
  year    = {2026},
  eprint  = {2605.18866},
  archivePrefix = {arXiv}
}

@article{weller1998tensorial,
  title   = {A tensorial approach to computational continuum mechanics using object-oriented techniques},
  author  = {Weller, Henry G. and Tabor, Gavin and Jasak, Hrvoje and Fureby, Christer},
  journal = {Comput. Phys.},
  volume  = {12},
  number  = {6},
  pages   = {620--631},
  year    = {1998},
  doi     = {10.1063/1.168744}
}

@misc{openfoam9,
  title        = {{OpenFOAM}: The Open Source {CFD} Toolbox, Version 9},
  author       = {{The OpenFOAM Foundation}},
  year         = {2021},
  howpublished = {Software},
  url          = {https://openfoam.org/version/9/}
}

@article{wilcoxon1945individual,
  title   = {Individual Comparisons by Ranking Methods},
  author  = {Wilcoxon, Frank},
  journal = {Biom. Bull.},
  volume  = {1},
  number  = {6},
  pages   = {80--83},
  year    = {1945},
  doi     = {10.2307/3001968}
}

@article{adrian1988stochastic,
  author  = {Adrian, R. J. and Moin, P.},
  title   = {Stochastic estimation of organized turbulent structure: homogeneous shear flow},
  journal = {J. Fluid Mech.},
  volume  = {190},
  pages   = {531--559},
  year    = {1988}
}

@inproceedings{zaheer2017deep,
  author    = {Zaheer, Manzil and Kottur, Satwik and Ravanbakhsh, Siamak and P{\'o}czos, Barnab{\'a}s and Salakhutdinov, Ruslan and Smola, Alexander J.},
  title     = {Deep Sets},
  booktitle = {Advances in Neural Information Processing Systems},
  volume    = {30},
  pages     = {3391--3401},
  year      = {2017}
}

@misc{prasthofer2022variable,
  author = {Prasthofer, Michael and De Ryck, Tim and Mishra, Siddhartha},
  title  = {Variable-Input Deep Operator Networks},
  year   = {2022},
  eprint = {2205.11404},
  archivePrefix = {arXiv}
}

@inproceedings{dulny2023dynabench,
  author    = {Dulny, Andrzej and Hotho, Andreas and Krause, Anna},
  title     = {{DynaBench}: A benchmark dataset for learning dynamical systems from low-resolution data},
  booktitle = {Machine Learning and Knowledge Discovery in Databases: Research Track ({ECML} {PKDD} 2023)},
  pages     = {438--455},
  publisher = {Springer},
  year      = {2023},
  doi       = {10.1007/978-3-031-43412-9_26},
  series    = {Lecture Notes in Computer Science},
  volume    = {14169},
  address   = {Cham}
}

@inproceedings{johnson2023oceanbench,
  author    = {Johnson, J. Emmanuel and Febvre, Quentin and Gorbunova, Anastasia and Metref, Sammy and Ballarotta, Maxime and Le Sommer, Julien and Fablet, Ronan},
  title     = {{OceanBench}: The sea surface height edition},
  booktitle = {Advances in Neural Information Processing Systems},
  volume    = {36},
  year      = {2023},
  pages     = {78275--78295},
  doi       = {10.52202/075280-3422}
}

@misc{pfister2026stocbench,
  title         = {{StocBench}: A Benchmark for Generative Modeling of Stochastic Dynamics},
  author        = {Sebastian Pfister and Benjamin Holzschuh and Nils Thuerey},
  year          = {2026},
  eprint        = {2608.22309},
  archiveprefix = {arXiv},
  url           = {https://arxiv.org/abs/2608.22309}
}

\end{document}